\documentclass[preprint, 
 amsmath,amssymb,
 aps, physrev,
]{revtex4-2}

\usepackage{enumitem}
\usepackage{graphicx}
\usepackage{dcolumn}
\usepackage{bm}
\usepackage{float}
\usepackage{booktabs} 
\usepackage{tabularx} 
\usepackage{mdframed}
\usepackage{xcolor}
\usepackage{changepage} 

\usepackage{booktabs}
\usepackage{array}
\usepackage{siunitx}
\usepackage{longtable}
\begin{document}

\preprint{arXiv Preprint}

\title{\textbf{Three Months in the Life of Cloud Quantum Computing} 
}%

\author{Darrell Teegarden}
 \email{Contact author: dteegard@umd.edu}
\author{Allison Casey}%
\author{F. Gino Serpa}%
\author{Patrick Becker}%
\author{Asmita Brahme }%
\author{Saanvi Kataria}%
\author{Paul Lopata}

\affiliation{%
 Authors' affiliations\\
  Applied Research Laboratory for Intelligence and Security (ARLIS), University of Maryland, College Park, MD
}%

\date{January 13, 2026}

\begin{abstract}
Quantum Computing (QC) has evolved from a few custom quantum computers, which were only accessible to their creators, to an array of commercial quantum computers that can be accessed on the cloud by anyone. Accessing these cloud quantum computers requires a complex chain of tools that facilitate connecting, programming, simulating algorithms, estimating resources, submitting quantum computing jobs, retrieving results, and more. Some steps in the chain are hardware dependent and subject to change as both hardware and software tools, such as available gate sets and optimizing compilers, evolve.

Understanding the trade-offs inherent in this process is essential for evaluating the power and utility of quantum computers. ARLIS has been systematically investigating these environments to understand these complexities. The work presented here is a detailed summary of three months of using such quantum programming environments. We show metadata obtained from these environments, including the connection metrics to the different services, the execution of algorithms, the testing of the effects of varying the number of qubits, comparisons to simulations, execution times, and cost. Our objective is to provide concrete data and insights for those who are exploring the potential of quantum computing.

It is not our objective to present any new algorithms or optimize performance on any particular machine or cloud platform; rather, this work is focused on providing a consistent view of a single algorithm executed using out-of-the-box settings and tools across machines, cloud platforms, and time. We present insights only available from these carefully curated data.
\end{abstract}

\maketitle


\section{\label{sec:level1}Introduction}

\subsection{\label{sec:level2}Motivation}
Quantum computing has emerged as a transformative technology that promises to revolutionize fields ranging from cryptography to materials science. However, benchmarking the practical utility and resource requirements of future quantum computers remains a subject of ongoing research \cite{Lubinski_2023, lubinski2024optimizationapplicationsquantumperformance, lorenz2025systematicbenchmarkingquantumcomputers}. Unlike efforts to rigorously quantify these aspects from an information theory perspective, this work focuses on documenting the usability of cloud quantum computers, using default settings, from a customer perspective. A variety of customers currently use cloud quantum computers to explore future applications, understand different platforms, or simply for workforce development. We anticipate that this cloud-access model will continue in the short- and mid-term as quantum computers will likely grow in size, cost, and complexity as they are scaled up.

This work aims to inform current and potential users of cloud quantum computers by providing user-focused metrics of multiple quantum computing service providers. Instead of focusing solely on rigorously optimized fidelity metrics, we present extensive data on machine accessibility, job submission times and success rates, performance in representative quantum circuits, and cost. We practically demonstrate how circuit transpiling methods significantly impact the outcomes of quantum circuits.

We hope that these user-centric measurements will help readers avoid the learning curve when engaging cloud quantum computing resources, maximize the value of submitted jobs, and ultimately save time and money. By providing detailed insights into the operational aspects of cloud quantum computing, this study serves as a valuable resource for both novice and experienced users. Our analysis encompasses a broad spectrum of quantum computing platforms, highlighting the strengths and limitations inherent in each. Through comprehensive data collection, our aim is to demystify the complexities associated with quantum job execution, offering clarity on performance variability and cost implications.

The data collected and analyzed in this study is intended to empower users with actionable information, enabling informed decision-making when selecting quantum service providers and optimizing quantum workloads. We emphasize the importance of understanding how different platforms handle quantum circuits and the resultant impact on execution fidelity and efficiency. Moreover, our findings underscore the significance of strategic planning in quantum computing endeavors, especially in terms of resource allocation and cost management.

Ultimately, this work advocates for a user-centric approach to cloud quantum computing, where accessibility, transparency, and practicality are prioritized. By fostering a deeper understanding of the quantum computing landscape, we aim to accelerate the adoption and integration of quantum technologies across diverse industries, paving the way for innovative applications and advancements.
\subsection{\label{sec:level2}Quantum computing on the cloud}
This study focused on the quantum resources available in the cloud from Microsoft (Azure Quantum) and Amazon Web Services (AWS Braket).
\subsubsection{\label{sec:level3}Microsoft Azure Quantum}
Microsoft Azure Quantum\cite{AzureQuantum} is a major provider of cloud quantum computing resources. These quantum resources are integrated with the Microsoft Azure classical cloud and high-performance computing resources, providing a highly capable framework for wide spectrum of high-performance computing.

The machines targeted for this study, available via Azure Quantum, are shown in Table~\ref{table:Azure_quantum_targets}

\begin{table}[h!]
\caption{Quantum Computers Available on Azure Quantum}
\centering
\renewcommand{\arraystretch}{1.25}
\setlength{\tabcolsep}{7pt}
\begin{tabularx}{\textwidth}{>{\bfseries}l l l c X}
\toprule
Vendor & Technology & Target Name & Qubits & Notes \\
\midrule
IonQ        & Trapped-ion     & Harmony & 11  & \cite{IonQHarmony2019}. Retired Sept 1, 2024 \\
            &                 & Aria-1  & 25  & \cite{IonQAria1} \\
            &                 & Aria-2  & 25  & \cite{IonQAria2}. Unavailable (maintenance) \\
            &                 & Forte   & 36  & \cite{IonQForte}. Private preview \\
\midrule
Quantinuum  & Trapped-ion     & H1-1    & 20  & \cite{QuantinuumH1} \\
            &                 & H2-1    & 56  & \cite{QuantinuumH2} \\
\midrule
Rigetti     & Superconducting & Ankaa-2 & 84  & \cite{RigettiAnkaa2}. Retired Oct 4, 2024 \\
\bottomrule
\end{tabularx}
\label{table:Azure_quantum_targets}
\end{table}

\subsubsection{AWS Braket}
The machines targeted for this study, available via AWS Braket\cite{AWSQuantum}, are shown in Table~\ref{table:AWS_quantum_targets}

\begin{table}[h!]
\caption{Quantum Computers Available on AWS Braket}
\centering
\renewcommand{\arraystretch}{1.25}
\setlength{\tabcolsep}{7pt}
\begin{tabularx}{\textwidth}{>{\bfseries}l l l c X}
\toprule
Vendor     & Technology      & Target Name & Qubits & Notes \\
\midrule
IonQ       & Trapped-ion     & Harmony     & 11     & Retired Sept 1, 2024 \\
           &                 & Aria-1      & 25     & \\
           &                 & Aria-2      & 25     & Unavailable (maintenance) \\
           &                 & Forte       & 36     & General use since Nov 22, 2024 \\
\midrule
IQM        & Superconducting & Garnet      & 20     & \cite{IQMGarnet} \\
\midrule
QuEra      & Neutral Atom    & Aquila      & 256    & \cite{QuEraAquila}. Analog, not gate-based, excluded from Qiskit Braket Provider \\
\midrule
Rigetti    & Superconducting & Ankaa-2     & 84     & Retired Oct 4, 2024 \\
\bottomrule
\end{tabularx}
\label{table:AWS_quantum_targets}
\end{table}

\subsection{\label{sec:level2}Structure}
The structure of this paper is as follows:
Section II outlines the methods and approach used in this study, including the development environment, backend, and workflow required to engage with quantum circuits on the cloud. Simulation of quantum circuits are discussed as well as the benchmark circuits commonly used for evaluating quantum machines. We also describe the strategy we used to collect meaningful data over time.

Section III provides results data and accompanying analysis. This includes information on queues for quantum machines in the cloud, availability, and job status. 
A primary result from this work is data and analysis on the fidelity of quantum circuit results as a function of several variables, including the number of circuit qubits, machine type, and calendar date. Measured fidelity is also compared to simulated fidelity, and a brief discussion of error models is presented.

Cost analysis is another important result of this work. The differing cost structure of the Microsoft and AWS platforms is described, and the resulting cost per machine type and as a function of circuit size is provided.

Section IV provides a discussion and conclusions, and Section V outlines future work.

\section{Methods and approach}

\subsection{Configuring a Development Environment}

Azure and AWS both provide an out-of-the-box cloud environment to utilize their respective quantum compute resources. Many practitioners, however, prefer to use their own, familiar development environment for executing quantum circuits on quantum hardware. This typically includes an Integrated Development Environment (IDE) such as Microsoft Visual Studio Code\cite{vscode}, Jupyter notebooks\cite{Jupyter}, and language-specific quantum libraries. 

Fortunately, configuring such a development environment is comparable across all platforms. Several quantum software development kits (SDK)/programming languages are available: Qiskit, Cirq, Braket, OCEAN, Q\#, etc. This study focused on IBM Open Source Python-Based Qiskit\cite{Qiskit} SDK since it supports numerous hardware vendors, allowing code reuse, and is widely used throughout the quantum community. Setting up a development environment is simply a matter of creating a python\cite{Python} virtual environment, installing the appropriate python packages, and setting the required environment variables.

\subsection{Job and Backend Monitoring}

Both AWS Braket and Microsoft Azure Quantum cloud services provide access to backend and job information via a web portal. Azure does not provide insight into the size or position of the job queue, while AWS provides limited information. Neither provides an accurate assessment of when a specific job will run, making it challenging to manage workflows and predict when results will become available. This will be discussed further in the Queue Wait Times and Machine Availability section, below. Azure provides the cost per job via the console and in the job data. The AWS Quantum Task console, however, does not display the cost per job; it can be calculated (see Cost section below) extracted from cost data for the Braket service in the AWS Cost and Billing Management Console, or estimated programmatically via the Bracket SDK Tracker module.

\subsection{Workflow}

The typical workflow for executing a quantum algorithm on a quantum computer is as follows:

\begin{enumerate}
    \item Identify (or create) a quantum algorithm, written in Python, using a quantum development kit such as Qiskit. This is called a quantum circuit. A quantum circuit defines the qubits in the system and the quantum gates that will operate sequentially on and between the qubits.
    \item Simulate the algorithm by executing the Python code on a classical computer. The simulation model typically assumes perfect qubit fidelity, allowing determination of the ideal, error-free result. This step is used to debug the circuit without the cost and delays associated with using a real quantum computer.
    \item (Optional) Simulate the algorithm using a classical computer, but with a simulation model that approximates the expected fidelity errors of the target quantum computer. This allows prediction of real-world results and validation of fidelity expectations for the target quantum computer.
    \item Execute the algorithm on a specific real quantum computer. This can be performed by updating the code to a specific \textquotesingle provider\textquotesingle{} and \textquotesingle backend\textquotesingle{} hardware to use for execution. The algorithm \textquotesingle run\textquotesingle{} is typically repeated many times (for example, 500 times, or \textquotesingle shots\textquotesingle{}), and the distribution of the resulting quantum states is tabulated. Multiple shots are necessary because of the natural distributed nature of quantum-qubit measurements.
    \item (Optional) Optimize quantum computer performance by specifying hardware-specific options that affect the transpiler (and its impact on the inputs to the quantum computer) and/or the post-processing of results. These options can significantly impact the fidelity of the results, the size of the circuit that can be submitted to the quantum computer, and the speed of execution. The use of these options is highly dependent on the specific hardware capabilities and can require deep knowledge of the specific system to be used effectively. \textbf{Please note that this work made no attempt to optimize performance using such options, but rather used the "out-of-the-box" settings for each machine. Consequently, these results should only be used to compare default performance, today, and not current and future potential performance as tuned by experts.}
    \item Analyze the results by examining the distribution of qubit values from the execution shots. It is often instructive to compare the resulting distributions with the ideal error-free simulation of the same circuit. This comparison allows calculation of the overall algorithm fidelity.
\end{enumerate}


\subsection{Simulation of quantum computer algorithms}

Quantum simulators (which use classical computers to emulate quantum computers, including quantum state evolution) have negligible queue wait times. Quantum circuits should be tested on simulators, prior to execution on real hardware, to minimize long wait times and cost. Simulated execution is possible when the total number of qubits in the circuit is small enough (generally, below 53 qubits). Although estimation tricks exist for classically simulating certain quantum circuits, it is generally considered impractical (due to memory constraints, available processing power, or cost) to simulate quantum circuits for more than 50 qubits using classical computing resources.

\subsection{Quantum Benchmarks for Cloud Quantum Assessment}

This study considered available quantum algorithm benchmarks as vehicles for assessing the available cloud quantum resources. Useful benchmark circuits are available in open source repositories; these two sources were evaluated:

\begin{enumerate}
    \item The Quantum Economic Development Consortium (QED-C) \cite{QEDC}
    \item The Qiskit circuit library \cite{QiskitLib}
\end{enumerate}

This study focused on the Quantum Fourier Transform (QFT) algorithm from the Qiskit circuit library for long-term testing and data collection. 

 See below (Qiskit library QFT algorithm benchmark) for more details.

\subsection{Cloud Quantum Resource Long-term Testing}

An important aspect of this testing work was to collect a consistent set of data across multiple cloud service providers (CSP), accessing quantum computers from multiple vendors, over a period that would allow meaningful analysis of the CSP and the vendor machines:

\begin{itemize}
    \item Quantum computer performance over time, algorithm size, vendor, and CSP
    \item Quantum computer availability and reliability over time, across vendor and CSP
    \item Quantum compute cost, comparing CSP, vendor, and algorithm size
\end{itemize}

\subsubsection{Cloud quantum data collection}

Data was collected by periodically running a quantum benchmark on various cloud quantum machines over slightly more than a 90-day time period, from mid-September 2024 through mid-December 2024. More than 5,000 jobs were submitted. The total cost for this dataset was approximately \$600,000.

 These quantum jobs had the following characteristics:

\begin{itemize}
    \item All jobs ran QFT and inverse QFT algorithms from the Qiskit circuit library.
    \item All jobs ran with default configurations; no additional optimizations were utilized.
    \item All jobs executed 500 shots.
    \item The QFT algorithm allows for a variable number of qubits; jobs were executed from 8 to 28 qubits, in 2-qubit increments (see the Qiskit library QFT algorithm benchmark below for more details).
    \item The jobs were run on both AWS Braket and Azure Quantum clouds.
    \item The jobs were executed on machines from IonQ (Aria 1, Aria 2, Forte 1), IQM (Garnet), and Quantinuum (H1, H2).
    \item Jobs were executed on simulators/emulators from IonQ (Aria, Forte) and Quantinuum (H1, H2).
\end{itemize}

\subsection{Data Collection and Analysis Software System}

\subsubsection{Data collection system requirements}

The following requirements were established when developing the software system for this data collection task:

\begin{itemize}
    \item The system must support at least Azure Quantum and AWS Braket as quantum CSPs.
    \item The system must automate daily job launch of at least one quantum algorithm on all available quantum machines.
    \item The system must manage launching jobs of a specified range of qubits for each available quantum machine.
    \item The system must periodically poll each CSP to discover and process available completed jobs.
    \item The system must automate processing of completed jobs to extract useful information (such as circuit fidelity, queue time, execution time, and cost) from the CSP.
    \item The system must save all job results such that all data can be analyzed without dependency on any CSP connection.
    \item The system must save normalized job data in a database that can be queried and filtered by any field.
    \item The system must provide a Python interface that allows direct queries of the database, returning DataFrame \cite{DataFrame} structures that allow for flexible analysis and visualization.
    \item The system should be extensible for adding additional CSPs and vendor machines.
    \item The system should be extensible for adding additional quantum algorithms, including specification of ideal results (for fidelity calculations).
\end{itemize}

\subsubsection{Data collection systems implementation}

Our team developed a software system that met all the above requirements. The system is written entirely in Python and utilizes a Mongo\cite{MongoDB} database hosted on an Azure virtual machine. The system was first prototyped by ARLIS student interns as part of the ARLIS RISC 2024 program\cite{RISC}. This ARLIS-developed data collection and analysis software normalizes the data across cloud providers and machine vendors, making visualizations consistent and simplifying analysis. Interfaces (abstract classes in Python) ensure the software is easily extendable, maintainable, consistent, and decoupled. The project source code is captured in a private GitHub repository.

\subsubsection{Qiskit library QFT algorithm benchmark}

As previously mentioned, testing focused on the Quantum Fourier Transform (QFT) algorithm. QFT was chosen for multiple reasons:

\begin{itemize}
    \item Key algorithm required to solve Shor’s algorithm.
    \item Circuit available in both the Qiskit library and QED-C benchmarks.
    \item Ideal result easy to determine.
\end{itemize}

Like the QED-C QFT benchmark, the circuit used for testing performs the following steps:

\begin{itemize}
    \item Initialize the qubits so that they represent the binary form of a randomly chosen number $n$. This is the number to which the QFT will be applied
    \item Perform the QFT.
    \item Perform a series of simple rotations to obtain the QFT applied to $n+1$
    \item Perform the inverse QFT.
\end{itemize}

The third bullet ("perform a series of rotations ...") merits some explanation. Had we omitted this step, in principle, we could just check if we recovered the original $n$ after performing the inverse. The problem with this is that the transpiler is always looking for simplifications to the quantum circuit and could, in principle, notice that we are performing a QFT and then an inverse QFT and create a null circuit. To avoid this, we take advantage of the fact that given the QFT of $n$ it is relatively easy to evaluate the QFT of $n+1$ by performing a series of simple rotations. So after adding the rotations $n+1$ is obtained after applying the inverse QFT.

\section{Data and analysis}

\subsection{Queue Wait Times and Machine Availability}

\subsubsection{Queue wait times}

Quantum computers are scarce resources. When a quantum job is submitted to a provider the job gets added to a queue. There may be a substantial wait time prior to the job executing. It is not clear when exactly a job that has been submitted will run. This may be for a variety of reasons:

\begin{enumerate}
    \item Hardware recalibrations may occur, delaying job processing.
    \item Vendor employees may be able to submit jobs ahead of the queue.
    \item Jobs have priorities associated with them (exact priority scheme may not be known).
    \item Jobs submitted via the “reservation” system take precedence over jobs in the queue at reservation time.
    \item Hardware may have planned or unplanned maintenance.
\end{enumerate}
 On the Azure platform, average queue time is the only available indicator as to when a job may run. Users do not have access to job position in the quantum provider’s queue or number of jobs in queue. 

The AWS platform does not provide an average queue time; instead, it provides limited queue information:

\begin{enumerate}
    \item Position of job in queue (see note below)
    \item Job priority (normal or priority)
    \item Number of jobs in “normal” queue
    \item Number of jobs in “priority” queue
\end{enumerate}
 Note: The queue information available in AWS is specifically for jobs submitted by AWS Braket customers, not an indicator of jobs submitted to the quantum device by all customers. This is important because the quantum computer vendors may share the same machines across multiple cloud providers or direct access.

\begin{figure}[H]
\includegraphics[width=\textwidth]{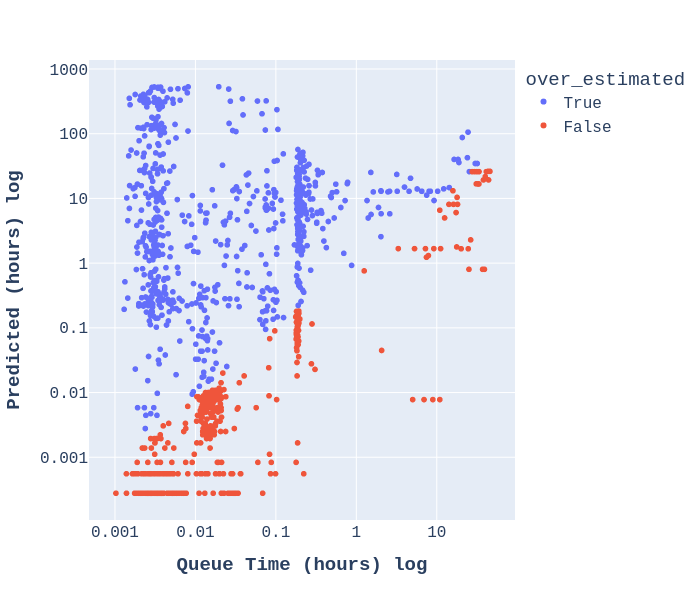}%
\caption{\label{fig:Queue_Time_Estimation}Predicted vs. actual job queue wait times (Azure). The corresponding data is not available for AWS.}
\end{figure}

Figure \ref{fig:Queue_Time_Estimation} shows a comparison of predicted versus actual queue wait times for the non-simulation jobs submitted via Azure. The blue data points represent jobs for which the vendor prediction of wait time was an overestimate. The red points represent an underestimate of wait queue time. If predicted queue times matched actual queue wait times, data would fall along a diagonal line from the bottom left to the upper right.

The data shows that the quantum machine vendors overestimated the queue time for 36\% of the jobs. The estimates, however, vary over such a large range that they provide relatively little value.

\subsubsection{Machine availability}

The figures below show the breakdown of the jobs submitted to the target machines, and the corresponding machine availability. Depending on the cloud provider and quantum vendor, a target will be in one of three possible states:
\begin{enumerate}
    \item\textbf{Available} – online, accepting jobs.
    \item\textbf{Degraded} – online, accepting jobs.
    \begin{enumerate}
        \item Definition of degraded varies by vendor. For Quantinuum, jobs are being accepted but not processed. Jobs will be processed once the system is back in the “Available” state.
        \item Rigetti, on the other hand, reported “Degraded” when the target was not operating with full qubit capacity.
    \end{enumerate}
    \item \textbf{Unavailable} – offline, not accepting job submissions.
    \begin{enumerate}
        \item For example, when a target is down for planned/unplanned maintenance. Periodically, machines are down for maintenance for a significant amount of time.
    \end{enumerate}
\end{enumerate}

\begin{figure}[H]
\includegraphics[width=\textwidth]{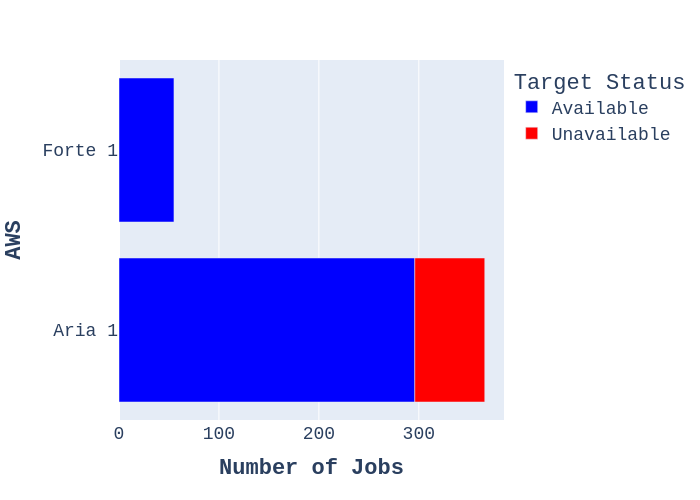}%
\caption{\label{fig:target_status_by_target_AWS}Target status by target on AWS}
\end{figure}

 Figure \ref{fig:target_status_by_target_AWS} shows the target status for jobs submitted on AWS.

\begin{figure}[H]
\includegraphics[width=\textwidth]{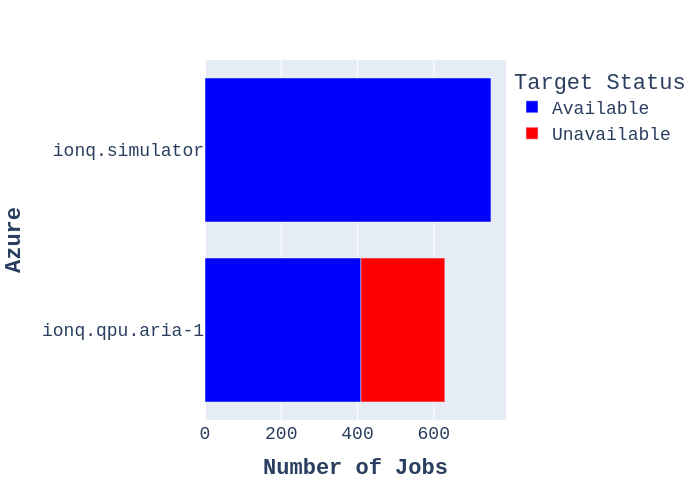}%
\caption{\label{fig:target_status_by_target_Azure}Target status for IonQ jobs submitted on Azure}
\end{figure}
 
Figure \ref{fig:target_status_by_target_Azure} shows the target status for jobs submitted in Azure, which all happen to be IonQ machines.

Note that ionq.simulator is a classical simulator, not a quantum computer.

\begin{figure}[H]
\includegraphics[width=\textwidth]{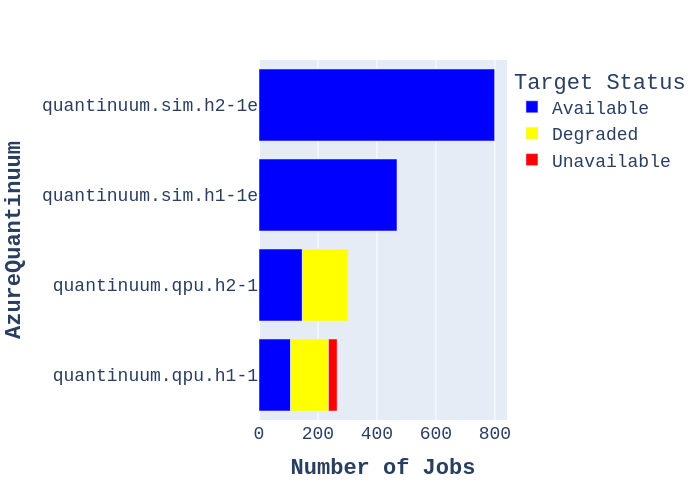}%
\caption{\label{fig:target_status_by_target_AzureQuantinuum}Target status for Quantinuum jobs submitted on Azure}
\end{figure}
Figure \ref{fig:target_status_by_target_AzureQuantinuum} shows the target status for Quantinuum jobs submitted on Azure.

Note that quantinuum.sim.h1(2)-1e are classical simulators, not quantum computers.

 \subsubsection{Target status over time}

We can view the availability, for each target, over the period that the data was collected. This shows us the availability profile for each machine. Note that this data is a snapshot of the availability only when attempting to submit a job. Consequently, gaps in any of the timeline plots below are simply due to lack of submitting a job or needing to allocate additional funding to a particular vendor before job submission could continue.

\begin{figure}[H]
\includegraphics[width=\textwidth]{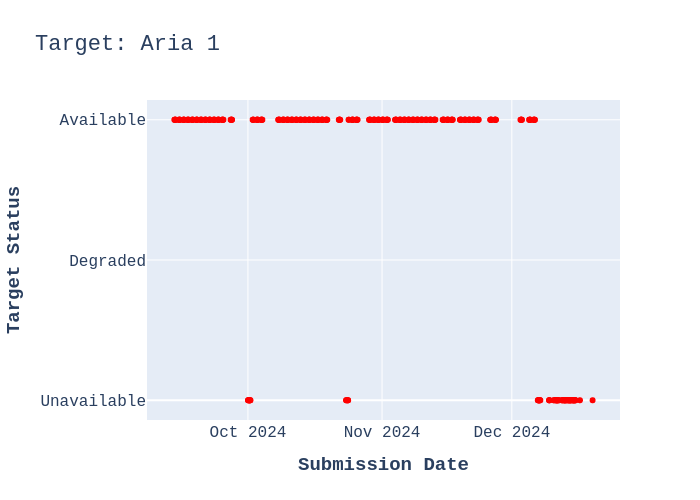}%
\caption{\label{fig:target_status_vs_time_Aria_1}Target status vs. time, Aria 1}
\end{figure}

Aria 1 had a good record of availability as shown in Figure \ref{fig:target_status_vs_time_Aria_1}. Aria 1 (AWS) was available approximately 81\% of the time. The availability of Aria 1 (Azure) was approximately 65\%. Twice the number of jobs were attempted for submission to Aria 1 (Azure), as opposed to Aria 1 (AWS), primarily, due to the AWS gate count difference (discussed in the Reliability and transparency issues section below). 


Aria 2 was unavailable, due to maintenance, for the entire long-term testing period. Aria 2, consequently, is omitted from the following analyses.

The Rigetti machine was frequently available, but often returned difficult-to-interpret error messages; the machine was retired from Azure in October 2024. The Rigetti machine, consequently, has been omitted from the analyses below.

 \begin{figure}[H]
\includegraphics[width=\textwidth]{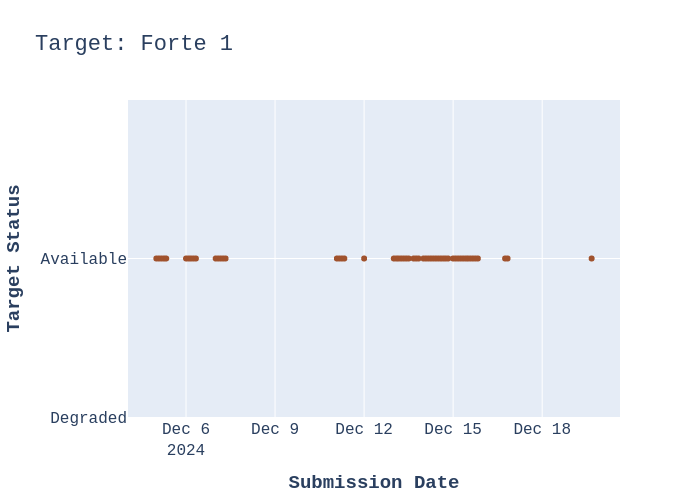}%
\caption{\label{fig:target_status_vs_time_Forte_1}Availability profile for the IonQ Forte 1 target}
\end{figure}

Figure \ref{fig:target_status_vs_time_Forte_1} shows that Forte 1 was available for submission on each attempt, keeping in mind it came online for general use on November 22, 2024. The ARLIS-developed software systems began submitting jobs to Forte 1 on December 5, 2024.

\begin{figure}[H]
\includegraphics[width=\textwidth]{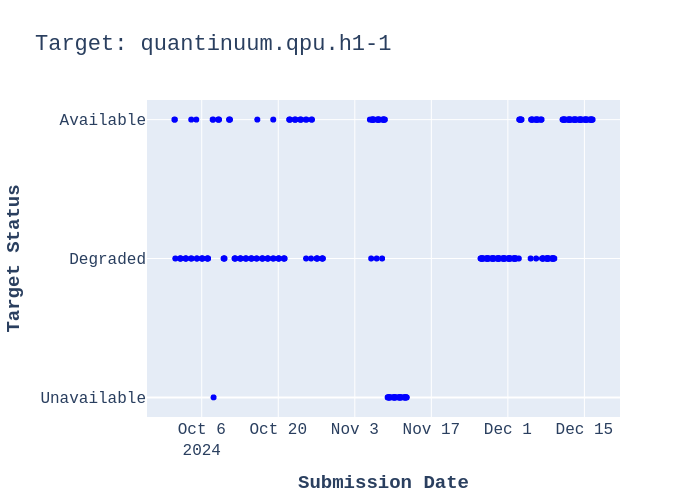}%
\caption{\label{fig:target_status_vs_time_quantinuum-h1}Availability profile for the Quantinuum h1-1 target}
\end{figure}

Quantinuum H1-1 frequently reported as degraded, as shown in Figure \ref{fig:target_status_vs_time_quantinuum-h1}. This is expected, as Quantinuum provides its customers with a monthly schedule indicating when their systems will be processing jobs (Available). Typically, this is between 5 PM – 2 AM Mountain Time on the days identified. Recall, jobs can be submitted while in the degraded state. H1-1 was available for job submission 89\% of the time.

\begin{figure}[H]
\includegraphics[width=\textwidth]{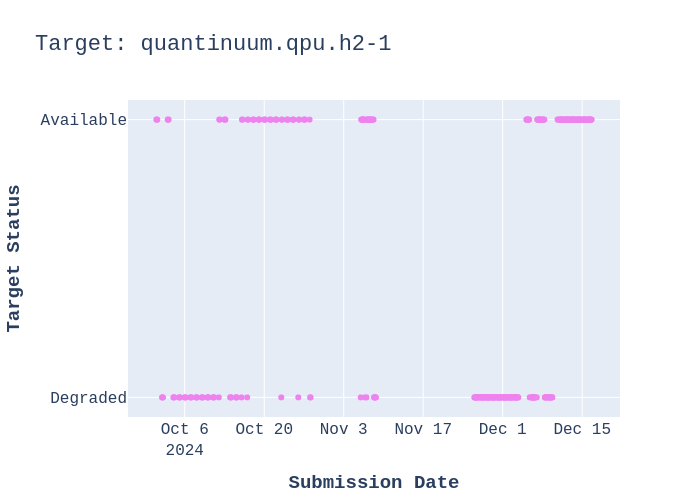}%
\caption{\label{fig:target_status_vs_time_quantinuum-h2}Availability profile for the Quantinuum h2-1 target}
\end{figure}

 Figure \ref{fig:target_status_vs_time_quantinuum-h2} shows that Quantinuum H2-1 was available 100\% of the time for job submission.

\subsubsection{Reliability and transparency issues}

ARLIS encountered challenges when using cloud providers to run on actual quantum hardware. The wait times in the queue prior to jobs executing was highly variable. Hardware could become unavailable with limited explanation. The mechanisms for managing subscriptions are not consistent across cloud providers and the quantum vendors supplying the hardware, resulting in confusion and occasional delays. Accessing the hardware via a third party introduces additional software and the potential for software issues that are not seen when accessing the hardware directly. For example, AWS rejected QFT circuits larger than 16 qubits due to exceeding gate count limits in AWS Braket. On Azure, QFT circuits up to 25 qubits were successfully submitted. It was determined that AWS was using a different gate set (assumed set of fundamental quantum gates) when transpiling for an IonQ backend, compared to either Azure or direct access. This discrepancy was reported to the hardware vendor. See Figure \ref{fig:fidelity_box} for more analysis of the impact of this difference.

\begin{figure}
    \centering
    \includegraphics[width=0.75\linewidth]{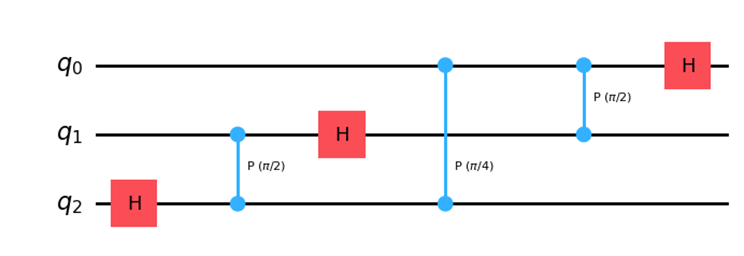}
    \caption{3 qubit QFT - Aria 1 – Azure}
    \label{fig:3_qubit_QFT_Aria_1_Azure}
\end{figure}

\begin{figure}
    \centering
    \includegraphics[width=1\linewidth]{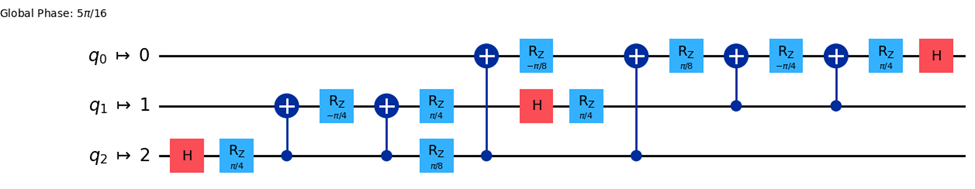}
    \caption{3 qubit QFT - Aria 1 - AWS Braket}
    \label{fig:3_qubit_QFT_Aria_1_AWS_Braket}
\end{figure}

Figures \ref{fig:3_qubit_QFT_Aria_1_Azure} and \ref{fig:3_qubit_QFT_Aria_1_AWS_Braket} illustrate the difference in transpiling the same 3 qubit QFT circuit for the IonQ Aria 1 machine via the Azure and AWS workflows. \textbf{The circuit transpiled using the Braket Qiskit provider (Figure \ref{fig:3_qubit_QFT_Aria_1_AWS_Braket}) is 3 times the size of same circuit transpiled using the Azure Qiskit provider (Figure \ref{fig:3_qubit_QFT_Aria_1_Azure})}.

\begin{adjustwidth}{1cm}{1cm}
\begin{mdframed}[backgroundcolor=lightgray, linecolor=black, nobreak=true]
\textbf{Operational Reliability Remains a Challenge}: Unpredictable queue times, limited transparency, and variable machine availability (e.g., Aria 2’s complete unavailability versus H2-1’s 100\% availability) highlight that cloud quantum computing is not yet a seamless, turn-key solution. 
\end{mdframed}
\end{adjustwidth}

\subsection{Job Status Analysis}

Jobs submitted via the ARLIS-developed system will be in one of five possible states:

\begin{itemize}
    \setlength{\itemsep}{0pt} 
    \setlength{\parskip}{0pt} 
    \item \textbf{Submitted} – Circuit accepted by target machine and awaiting execution or final processing.
    \item \textbf{Processed} – Circuit successfully completed execution and results calculated and stored.
    \item \textbf{Error} – Error submitting or executing job on target machine.
    \item \textbf{Canceled} – Job canceled before execution began.
    \item \textbf{Unavailable} – Job unable to be submitted, target system unavailable.
\end{itemize}

\subsubsection{Job status by target}

Prior to submitting a circuit to a backend, the circuit is transpiled for the target machine (the Qiskit code is mapped to a specific machine implementation). The ARLIS-developed system tracks such transpiler errors, as well as errors that occur on job submission, and saves them for later analysis. 

\begin{figure}[H]
\includegraphics[width=\textwidth]{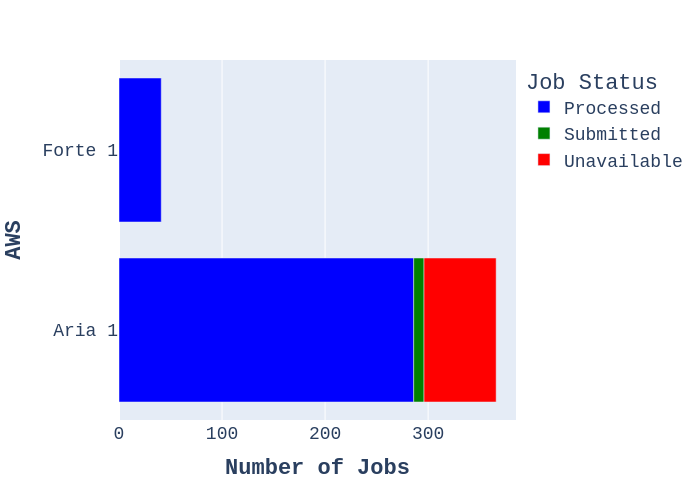}%
\caption{\label{fig:job_status_by_target_AWS}Machine availability on AWS Braket}
\end{figure}

\begin{figure}[H]
\includegraphics[width=\textwidth]{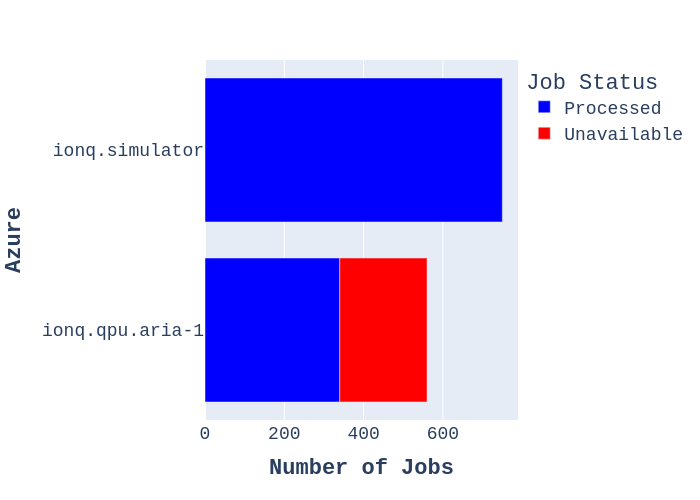}%
\caption{\label{fig:job_status_by_target_Azure}IonQ machine availability on Azure}
\end{figure}

\begin{figure}[H]
\includegraphics[width=\textwidth]{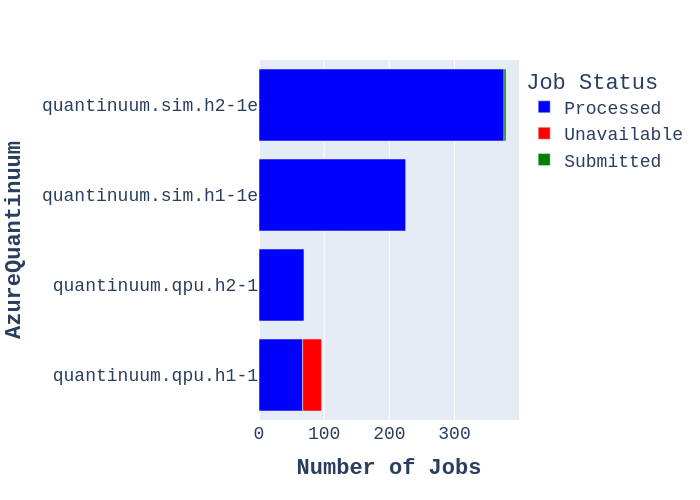}%
\caption{\label{fig:job_status_by_target_AzureQuantinuum} Quantinuum machine availability on Azure}
\end{figure}

 Generally, this study found few errors running jobs on IQM, IonQ and Quantinuum systems. The Forte machine produced errors due to the gate counts limits previously discussed. The Aria and Quantinuum machines produced errors due to reaching subscription quota limitations (see Cost section below). These errors have been omitted in Figures \ref{fig:job_status_by_target_AWS}, \ref{fig:job_status_by_target_Azure}, and \ref{fig:job_status_by_target_AzureQuantinuum} since they could have been avoided by better management of circuit submission and are not indicative of intrinsic machine limitations.

Figure \ref{fig:job_status_by_target_AWS} shows the status of the jobs on AWS for Garnet, Forte 1, and Aria 1 machines. The Aria 2 machine is not shown, as it was unavailable due to maintenance for the entire testing period.

\subsection{Target Machine Fidelity Analysis}

The ARLIS-developed system uses the Qiskit quantum information library to calculate Hellinger Fidelity. Hellinger Fidelity is a measure of similarity between two probability distributions.\cite{HellingerFidelity} The fidelity values range from zero to one, with one indicating identical distributions. A fidelity threshold of \begin{equation}
    \frac{1}{e} \approx 0.37 \label{eq:inverse_of_e}
\end{equation} is used as a natural fidelity cutoff value \cite{IonQForte}, when categorizing runs into bins of success and failure, based on fidelity results (fidelity above 0.37 is considered successful; fidelity below 0.37 is considered a failure). Observed circuit fidelity (\textit{F}) is expected to vary to first order exponentially with  gate count (n\textsubscript{gates}), and individual gate fidelity (\textit{F\textsubscript{gate} }).

\subsubsection{Fidelity as a function of number of qubits}

\begin{figure}[H]
\includegraphics[width=\textwidth]{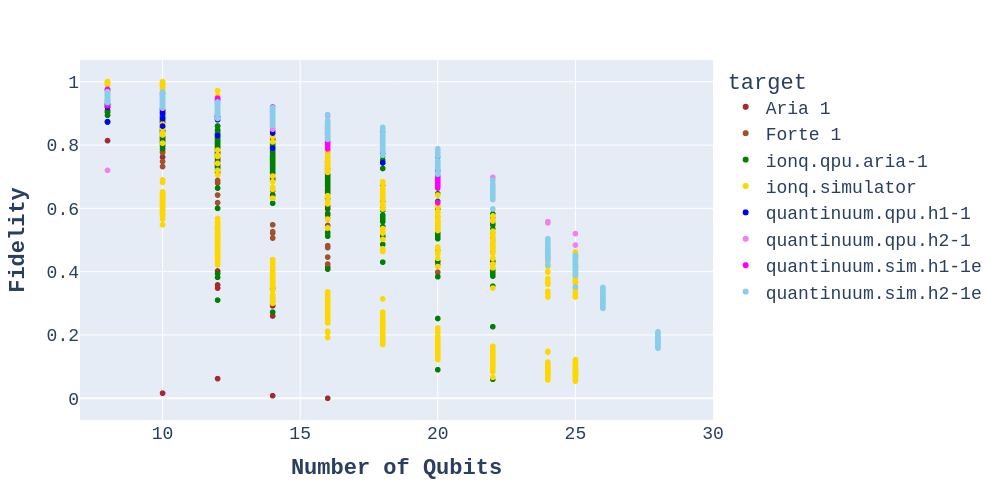}%
\caption{\label{fig:fidelity_vs_qubits}QFT fidelity as a function of number of qubits}
\end{figure}

 Figure \ref{fig:fidelity_vs_qubits} shows the fidelity measurements observed over the entire range of all the submitted jobs. As expected, the largest (best) values are near 1.0 and generally decrease with increasing number of qubits. Note, for example, the fidelity range for the 20-qubit QFT algorithm: it is as low as 0.0 and as high as 0.8. If these machines were operating perfectly, fidelity would deviate from 1 only due to shot noise (finite number of shots).

The range of fidelity values produced by the target quantum computers is high, indicating a wide range of performance of these machines

\begin{figure}[H]
\includegraphics[width=\textwidth]{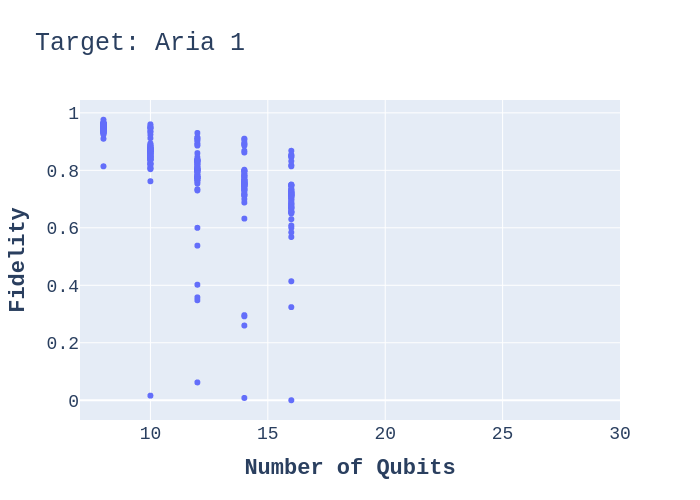}%
\caption{\label{fig:fidelity_vs_nqubits_Aria_1}IonQ Aria 1 (AWS) QFT fidelity as a function of number of qubits}
\end{figure}

\begin{figure}[H]
\includegraphics[width=\textwidth]{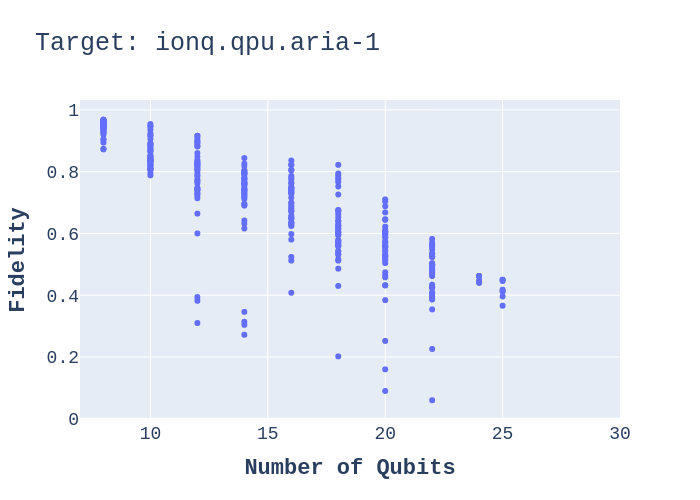}%
\caption{\label{fig:fidelity_vs_nqubits_ionq} IonQ Aria 1 (Azure) QFT fidelity as a function of number of qubits}
\end{figure}

Figure \ref{fig:fidelity_vs_nqubits_Aria_1} and Figure \ref{fig:fidelity_vs_nqubits_ionq} break out the QFT algorithm fidelity as a function of number of qubits for the Aria 1 machine from IonQ. These are results from the same machine, simply accessed via two different cloud providers (AWS and Azure). 

This illustrates an important error that we discovered: AWS was using the wrong quantum gate set when sending jobs to IonQ. This significantly reduces the qubit size of the algorithms that can be sent to IonQ from AWS!

As previously mentioned, this gate set discrepancy has been reported to IonQ staff.

\begin{adjustwidth}{1cm}{1cm}
\begin{mdframed}[backgroundcolor=lightgray, linecolor=black, nobreak=true]
\textbf{Transpiler Configuration Critically Impacts Performance}: AWS Braket’s suboptimal gate set for IonQ’s Aria 1 produced circuits larger than Azure’s, limiting qubit counts to 16 versus Azure’s 25. Users must consider transpiler impacts to maximize hardware utilization.
\end{mdframed}
\end{adjustwidth}

\begin{figure}[H]
\includegraphics[width=\textwidth]{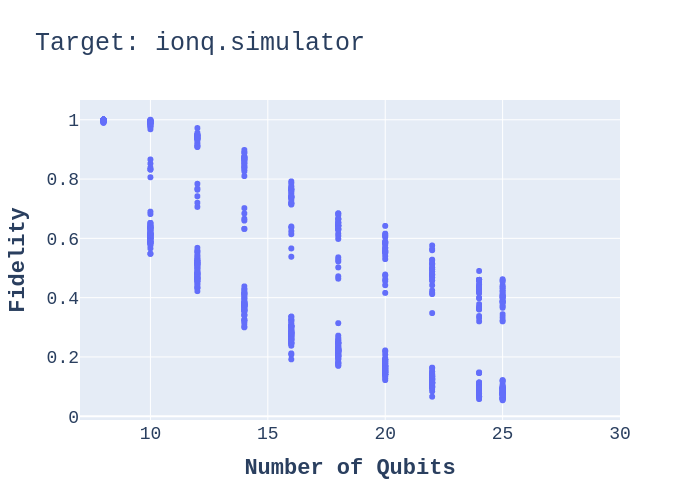}%
\caption{\label{fig:fidelity_vs_nqubits_ionq_simulator} IonQ Aria simulator QFT fidelity as a function of number of qubits}
\end{figure}

Figure \ref{fig:fidelity_vs_nqubits_ionq_simulator} shows the simulated fidelity of IonQ’s Aria 1 machine, as a function of the number of qubits.

Note the bimodal distribution of the simulation results. See the simulation results as a function of time in Figure \ref{fig:fidelity_vs_time_aria_simulator} to see how the simulations changed during the data collection period. 

It appears that IonQ staff modified the simulation parameters during the data collection period, presumably to improve the match between measurement and simulation.

\begin{figure}[H]
\includegraphics[width=\textwidth]{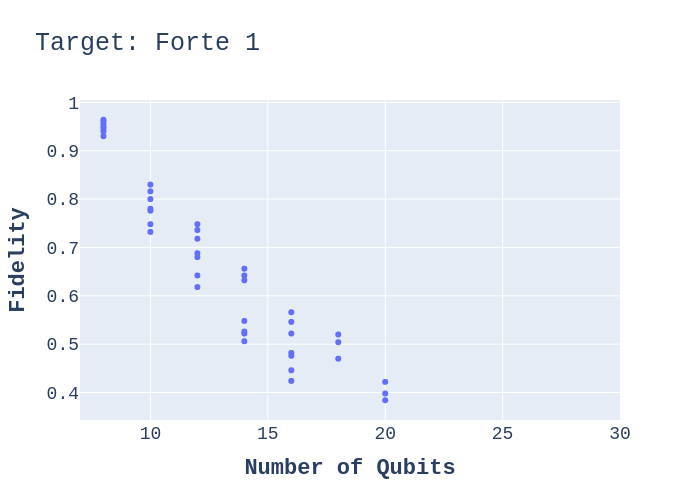}%
\caption{\label{fig:fidelity_vs_nqubits_Forte_1} IonQ Forte 1 (AWS) QFT fidelity as a function of number of qubits}
\end{figure}

\begin{figure}[H]
\includegraphics[width=\textwidth]{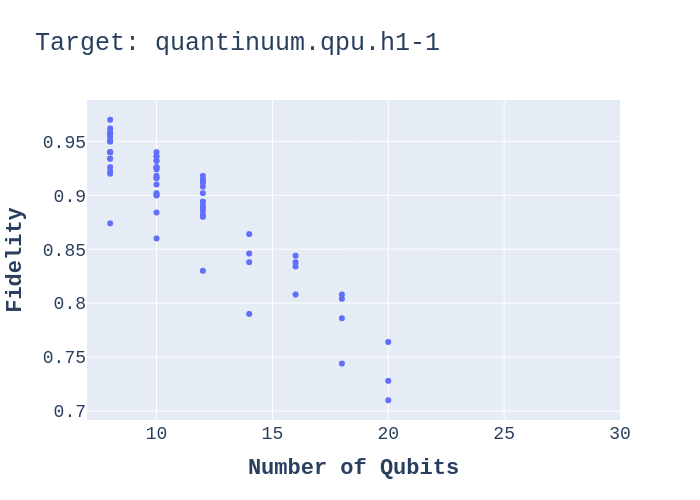}%
\caption{\label{fig:fidelity_vs_nqubits_quantinuum-h1}Quantinuum h1-1(Azure) QFT fidelity as a function of number of qubits}
\end{figure}

\begin{figure}[H]
\includegraphics[width=\textwidth]{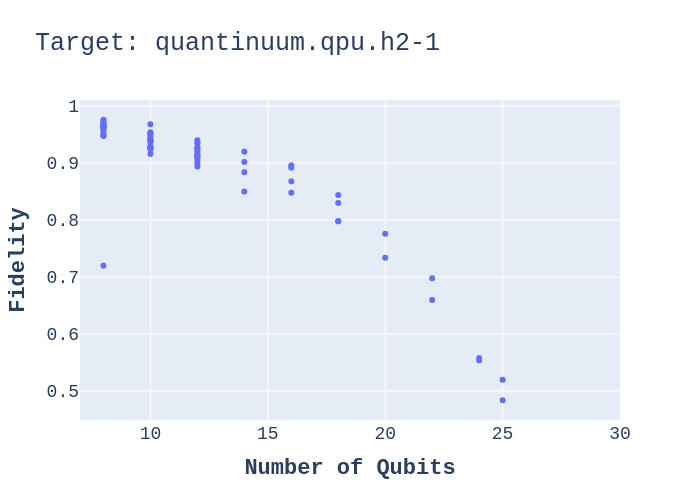}%
\caption{\label{fig:fidelity_vs_nqubits_quantinuum-h2}Quantinuum h2-1(Azure) QFT fidelity as a function of number of qubits}
\end{figure}

The following plots in Figures \ref{fig:fidelity_vs_nqubits_Forte_1}, \ref{fig:fidelity_vs_nqubits_quantinuum-h1}, and \ref{fig:fidelity_vs_nqubits_quantinuum-h2} show the breakout for fidelity versus number of qubits for the highest performance offerings from Ionq (Forte 1) and Quantinuum (H1-1 and H2-1). 

\subsubsection{Fidelity as a function of machine}

\begin{figure}[H]
\includegraphics[width=\textwidth]{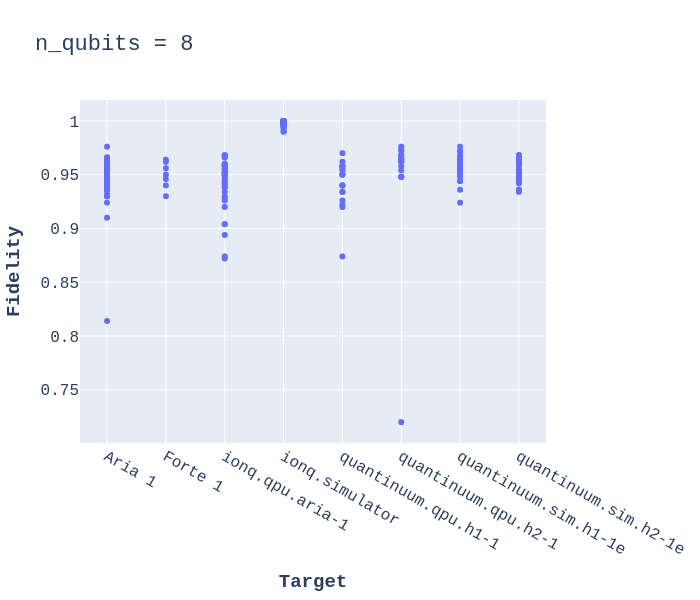}%
\caption{\label{fig:fidelity_vs_target_8}8-qubit QFT Fidelity as function of target}
\end{figure}

\begin{figure}[H]
\includegraphics[width=\textwidth]{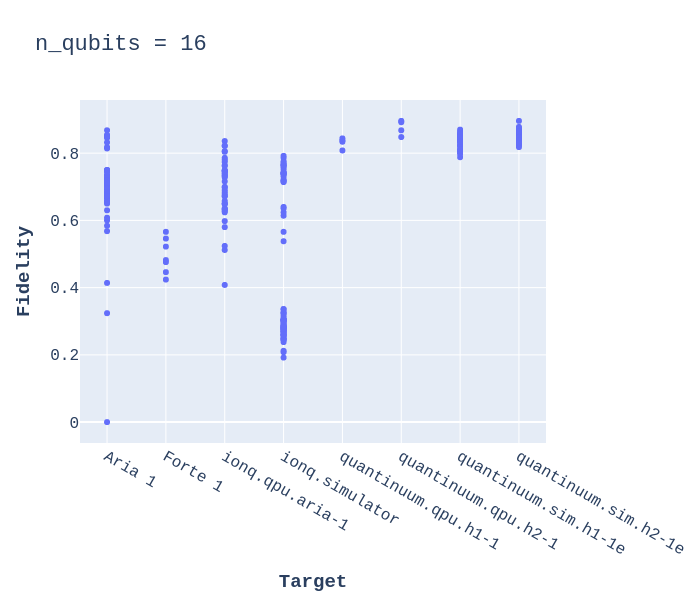}%
\caption{\label{fig:fidelity_vs_target_16}16-qubit QFT Fidelity as function of target}
\end{figure}

\begin{figure}[H]
\includegraphics[width=\textwidth]{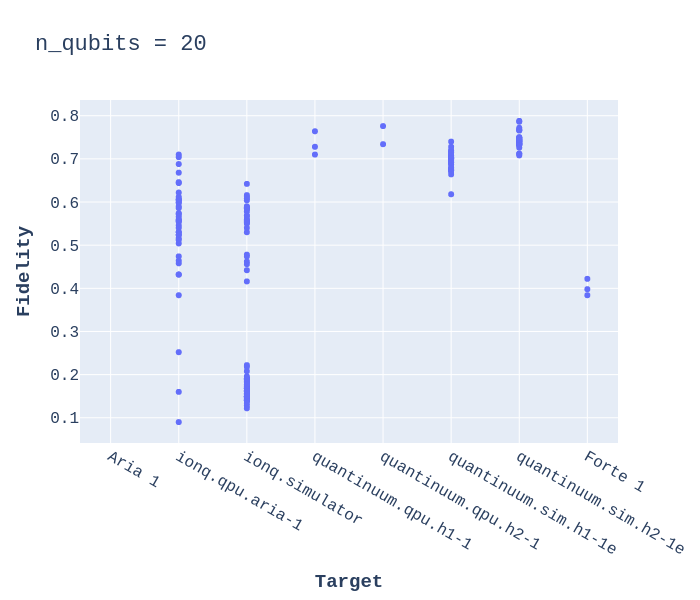}%
\caption{\label{fig:fidelity_vs_target_20}20-qubit QFT Fidelity as function of target}
\end{figure}

\begin{figure}[H]
\includegraphics[width=\textwidth]{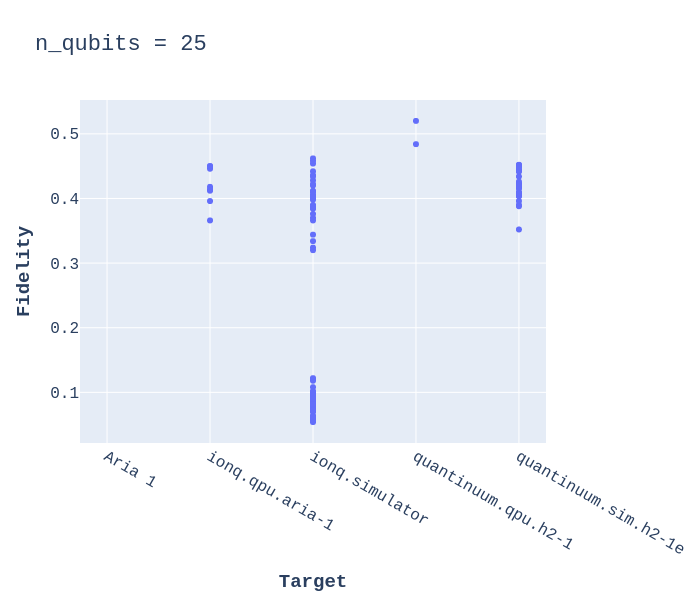}%
\caption{\label{fig:fidelity_vs_target_25}25-qubit QFT Fidelity as function of target}
\end{figure}

The plots in this section provide breakouts of the fidelity across machines, but for specific qubit counts. Representative qubit counts of 8, 16, 20, and 25 are provided in Figures \ref{fig:fidelity_vs_target_8}, \ref{fig:fidelity_vs_target_16}, \ref{fig:fidelity_vs_target_20}, and \ref{fig:fidelity_vs_target_25}, respectively.

The performance of the machines becomes more variable at higher qubit counts.

\subsubsection{Fidelity as a function of time}

\begin{figure}[H]
\includegraphics[width=\textwidth]{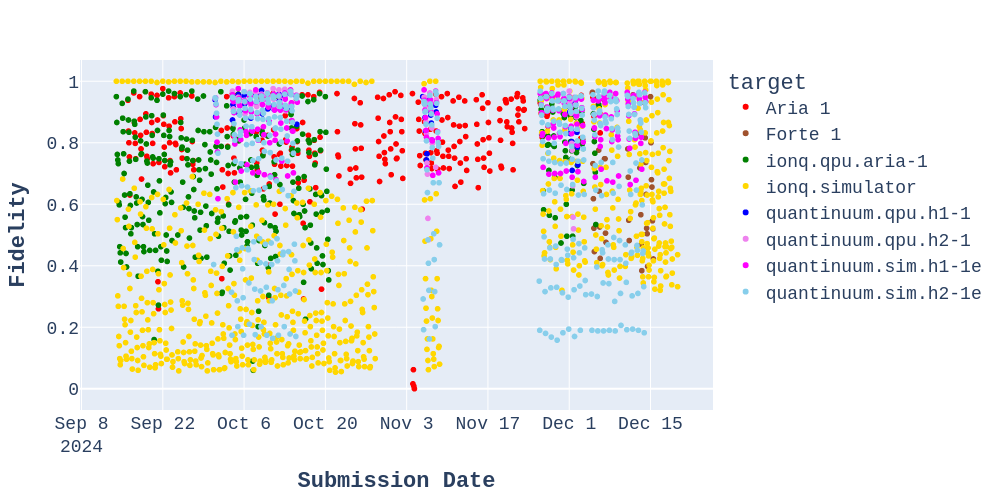}%
\caption{\label{fig:fidelity_vs_time}QFT fidelity over time for all targets and number of qubits}
\end{figure}

\begin{figure}[H]
\includegraphics[width=\textwidth]{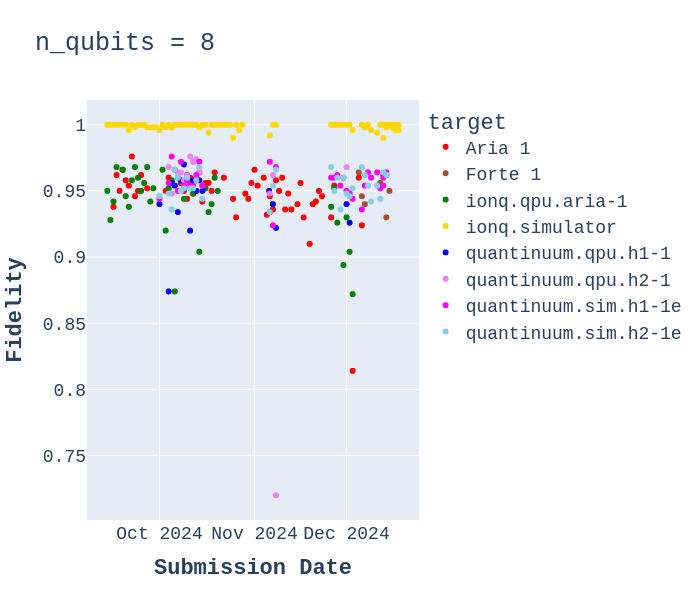}%
\caption{\label{fig:fidelity_vs_time_8}QFT fidelity over time for all targets and 8 qubits}
\end{figure}

\begin{figure}[H]
\includegraphics[width=\textwidth]{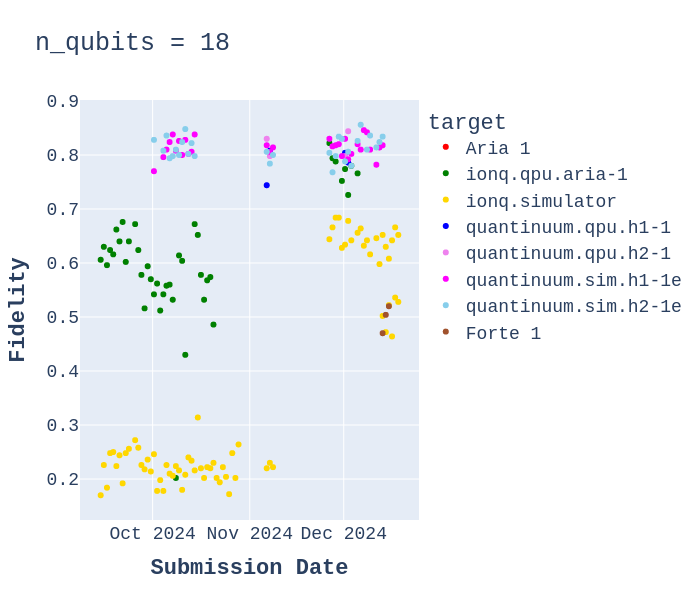}%
\caption{\label{fig:fidelity_vs_time_18}QFT fidelity over time for all targets and 18 qubits}
\end{figure}

\begin{figure}[H]
\includegraphics[width=\textwidth]{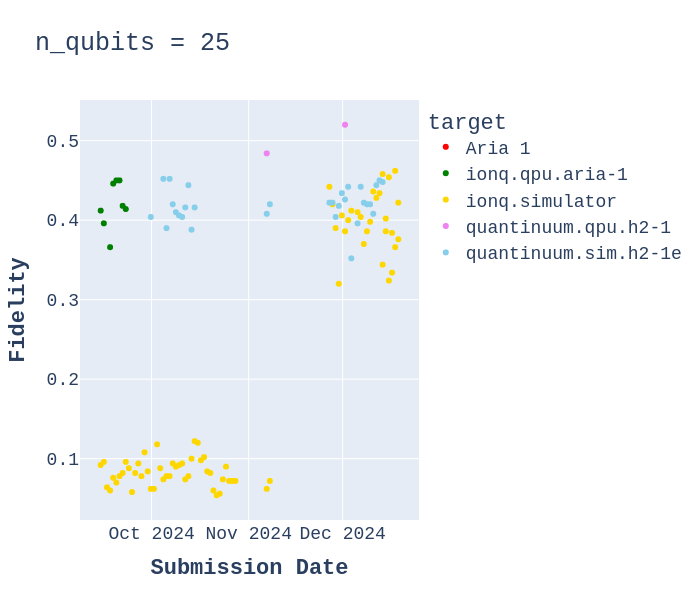}%
\caption{\label{fig:fidelity_vs_time_25}QFT fidelity over time for all targets and 25 qubits}
\end{figure}

Our data reveals that machine output fidelity varied over time. Figures \ref{fig:fidelity_vs_time}, \ref{fig:fidelity_vs_time_8}, \ref{fig:fidelity_vs_time_18}, and \ref{fig:fidelity_vs_time_25} plot this fidelity data, highlighting significant performance changes.

 The first plot shows the entire data set, including all machines and qubit sizes. The data density on this plot, however, is too high to gain much insight. Subsequent plots show breakouts for selected number of qubits (8,18, and 25). These plots provide better visibility into the stability of machine results over time.

\subsubsection{Measured fidelity compared to simulated fidelity}

\begin{figure}[H]
\includegraphics[width=\textwidth]{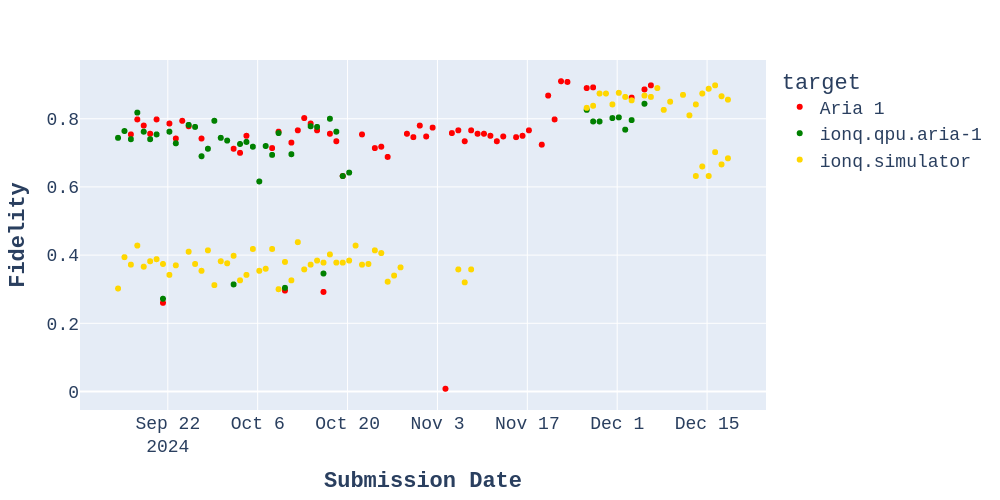}%
\caption{\label{fig:fidelity_vs_time_aria_simulator}Fidelity vs. submission time for IonQ Aria 1 (AWS), ionq.qpu.aria-1 (Azure) and ionq.simulator (Azure) for 14-qubit QFT}
\end{figure}

Figure \ref{fig:fidelity_vs_time_aria_simulator}, below, shows the time-varying fidelity for both the Aria-1 machine and the equivalent simulation. This shows how the simulator progresses from poor agreement to general agreement beginning in late November. Speculation about the causes (such as changing noise models) is possible, but a more detailed analysis would be required to provide a better explanation.

Before November 17 for Aria 1, the IonQ simulator underestimates the fidelity by as much as 80\%. After which we find closer agreement, ranging from 3\% to 18\%.

\begin{figure}[H]
\includegraphics[width=\textwidth]{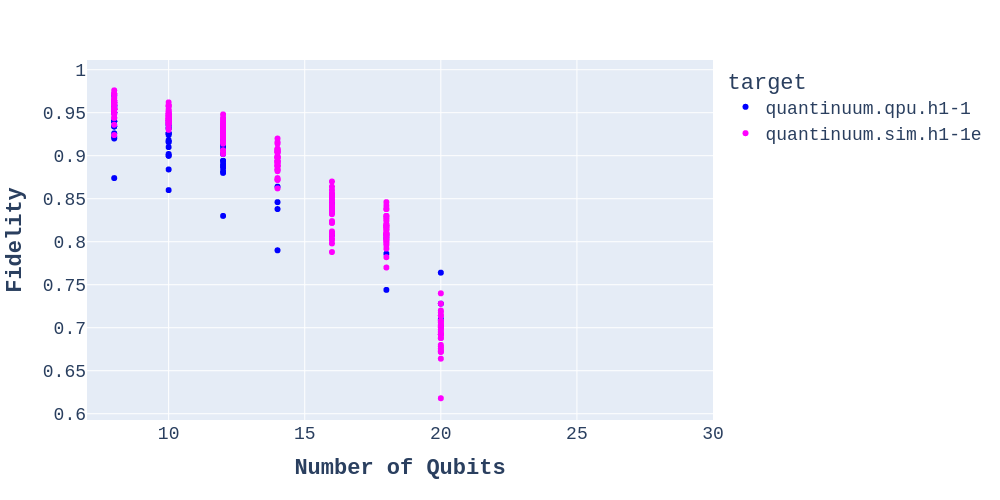}%
\caption{\label{fig:h1_vs_sim}Quantinuum H1-1 vs. Emulator}
\end{figure}

\begin{figure}[H]
\includegraphics[width=\textwidth]{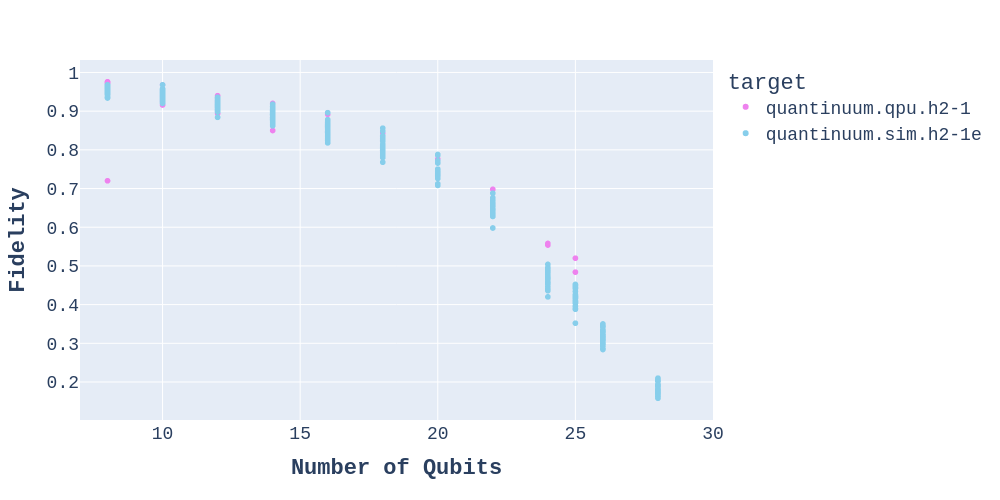}%
\caption{\label{fig:h2_vs_sim}Quantinuum H2-1 vs. Emulator}
\end{figure}

 Further details of Quantinuum’s H1 and H2 hardware versus their corresponding emulators are shown in Figure \ref{fig:h1_vs_sim} and Figure \ref{fig:h2_vs_sim}. For H1 the fidelities predicted by the Quantinuum emulator and the fidelities produced by actual hardware are within 5\% for 20 qubits and down to less than 2\% for 8 qubits. This trend is what one would expect because of the increase in complexity for larger number of qubits. Similarly, for H2 the fidelities predicted by the Quantinuum emulator and the fidelities produced by actual hardware are within 16\% for 25 qubits and down to less than 1 percent for 8 qubits. 

\subsubsection{Zero-order noise models for fidelity}

\begin{figure}[H]
\includegraphics[width=\textwidth]{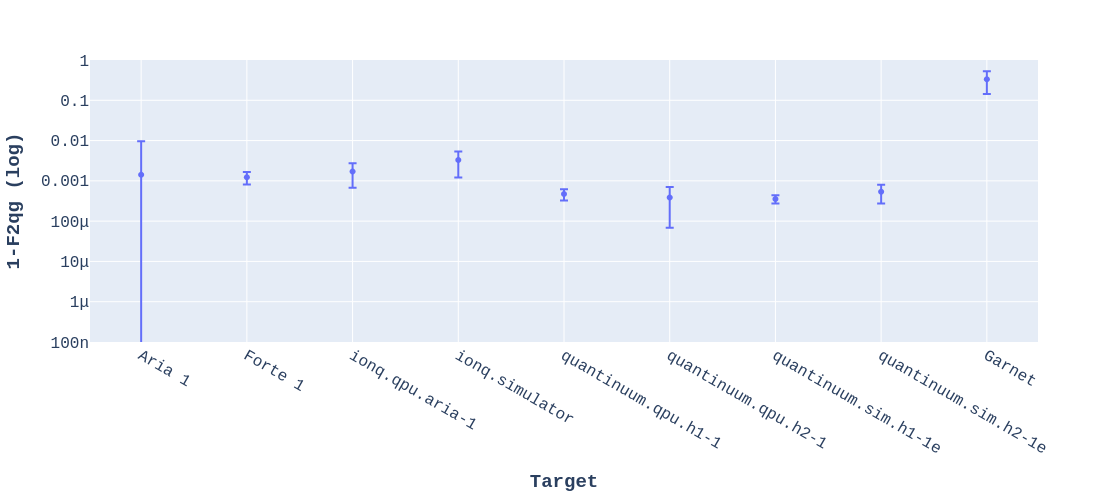}%
\caption{\label{fig:error_vs_target}Error rate (1-2-qubit gate fidelity) vs. target}
\end{figure}

Having run many jobs on different machines and under different conditions, it is natural to ask whether one can extract error rates and how they compare for different targets. Estimating fidelities from noise models and algorithms is difficult but using a simple model one can try to obtain some information from the data set. To a first approximation one can assume that most of the error occurs in the 2 qubit gates. We model the fidelity as the result of applying a series of 2 qubit gates each contributing to the noise with a factor $F_{2qg}$. Consequently, the overall fidelity, F, can be expressed as $F_{2qg}^{n}$   (where $n$ is the number of 2 qubit gates). This simple model allows inference of the value of $F_{2qg}$ for each run. The figure above summarizes the results obtained from applying this simple model. 

 Since the fidelity values of single 2 qubit gates is close to 1, the quantity 1-$F_{2qg}$ (the error) is reported in the y axis (in log scale); the QPU targets are shown on the x axis in Figure \ref{fig:error_vs_target}. Each dot is the average of the error for all the runs performed on a given target while the error bars are 1 standard deviation. The values obtained are consistent with a value for $F_{2qg}$ of approximately 0.99, which to first order seems plausible. 

\begin{figure}[H]
\includegraphics[width=\textwidth]{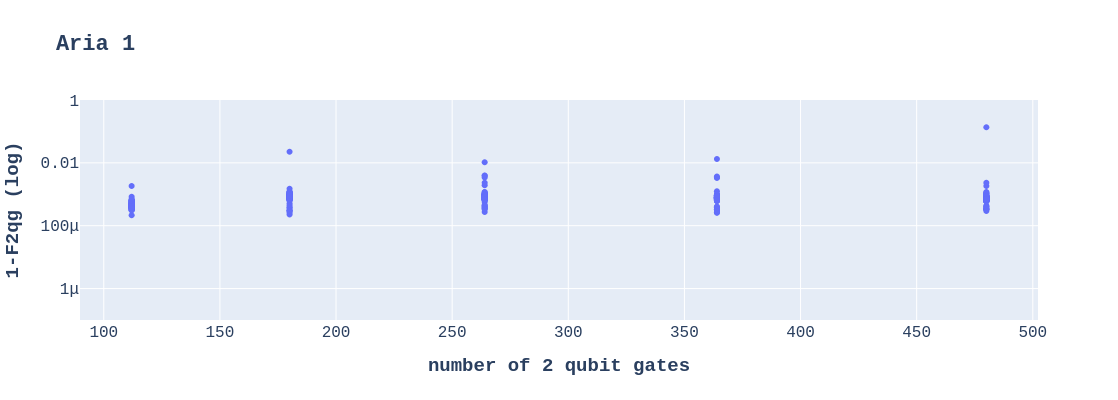}%
\caption{\label{fig:error_vs_2qg_arial}Two qubit gate error rates as a function of number of 2 qubit gates for Aria-1, 8 qubits}
\end{figure}

 All the error bars seem reasonable given the simple model. The only error bar that seems too large is the one corresponding to Aria-1. Figure \ref{fig:error_vs_2qg_arial}, below, shows the individual errors for Aria-1. The large error bar is explained by the outliers (top of the figure) that distort the standard deviation (recall this is in log scale). Despite this, the inferred errors for Aria1 (AWS) and the ionq-aria-1 (Azure) are consistent with each other. Figure \ref{fig:h1_vs_sim} and Figure \ref{fig:h2_vs_sim}, above, shows that the emulators for Quantinuum compare well to the hardware. A more detailed analysis of the Quantinuum noise models may be useful for comparing model parameters.

Including all runs in our error computations may introduce time-effects (such as machine calibration and changing noise models), leading to these conservative error bars.

\begin{figure}[H]
\includegraphics[width=\textwidth]{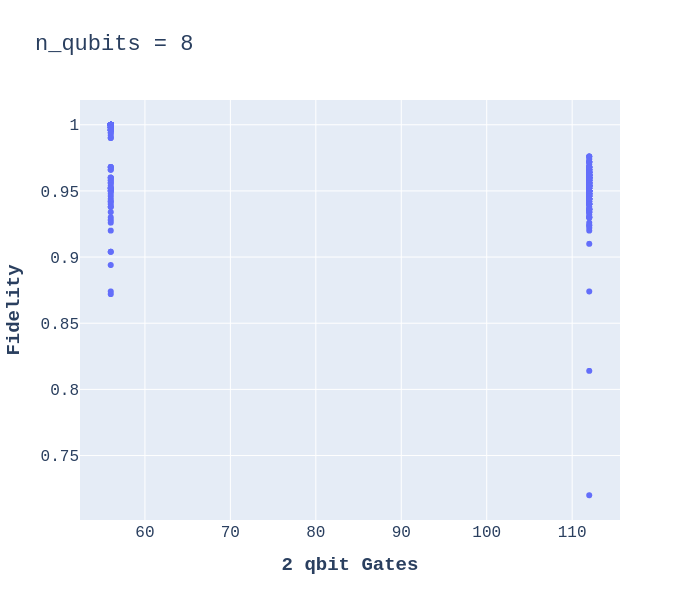}%
\caption{\label{fig:fidelity_vs_2qgates_8}8-qubit fidelity as a function of 2-qubit gate count}
\end{figure}

 The impact of the transpiler gate set can be seen in Figure \ref{fig:fidelity_vs_2qgates_8}. This shows the fidelity as a function of 2-qubit gates in the circuit for Aria 1, 8 qubits QFT. If the transpilers for Azure and AWS were both using the same target gate set, the number of 2-qubit gates for an 8-qubit QFT would all be the same. Here we see that the AWS transpiler produces a target circuit with approximately twice the 2-qubit gates as the Azure transpiler produces. Intuitively, one would expect the fidelity to be lower for the version with a larger number of 2-qubit gates. However, while the difference in fidelity between these two is apparent, it is not as large as one might expect.

\begin{figure}[H]
\includegraphics[width=\textwidth]{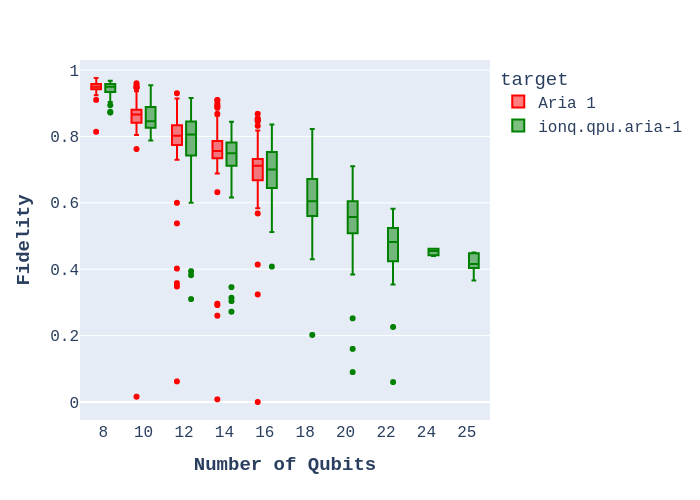}%
\caption{\label{fig:fidelity_box}Fidelity as a function of qubits, AWS Aria 1 (blue) and Azure ionq.qpu.aria-1 (red)}
\end{figure}

Figure \ref{fig:fidelity_box} provides a more detailed examination of this difference. It shows that the reported fidelity for AWS-submitted runs, despite requiring a larger number of two-qubit gates, is comparable to the fidelity reported for Azure. The notable distinction, however, lies in the greater qubit range offered by the Azure-submitted runs. It's important to remember that each data point in this plot starts with the same Qiskit circuit and targets the same quantum machine, Aria 1, located at the IonQ data center. The difference between the red and blue data points is solely attributable to the cloud provider and the tool chain used to submit the circuit to the quantum machine.

\subsubsection{Fidelity as a function of 2-qubit gates}

\begin{adjustwidth}{1cm}{1cm}
\begin{mdframed}[backgroundcolor=lightgray, linecolor=black, nobreak=true]
\textbf{Fidelity Varies Widely Across Machines and Time}: The significant range in QFT fidelity (0.0 to 0.8 for 20 qubits) and the initial inaccuracy of IonQ’s simulator reveal performance inconsistencies. Quantinuum’s close emulator-hardware alignment suggests more reliable simulation, critical for debugging and predicting real-world results.
\end{mdframed}
\end{adjustwidth}

\subsection{Cloud Quantum Cost Structure}

Cost varies depending upon the quantum hardware vendor and cloud service provider. 

\subsubsection{Azure IonQ Cost\cite{MSFTQuantumPricing}}

Azure Quantum comes with \$500 of free quantum credit per quantum provider after which pay-as-you-go and subscriptions options are available. The IonQ pay-as-you-go pricing plan for Azure is

Cost = (Total 1-qubit gate cost + Total 2-qubit gate cost) * Number of Shots

where:

Total 1-qubit gate cost = Number of 1-qubit gates * 1-qubit gate price 

Total 2-qubit gate cost = Number of 2-qubit gates * 2-qubit gate price

 Table \ref{tab:azure_pricing} provides the Azure cost breakdown for Aria on Azure. There are minimum per circuit costs depending upon error mitigation being enabled/disabled. These fixed costs dominate the overall cost for small circuit runs with error mitigation enabled.

\begin{table}[h!]
\caption{Azure On-Demand Pricing}
\centering
\renewcommand{\arraystretch}{1.3}
\begin{tabular}{l@{\hspace{0.3cm}}r@{\hspace{0.3cm}}r@{\hspace{0.3cm}}p{6cm}}
\hline
\textbf{Device} & \textbf{1-Qubit Gate Price} & \textbf{2-Qubit Gate Price} & \textbf{Notes} \\
\hline
Aria & \$0.00022 & \$0.000975 & \$12.42 minimum per circuit without error mitigation. \newline \$97.50 minimum per circuit with error mitigation enabled. \\
\hline
\end{tabular}
\label{tab:azure_pricing}
\end{table}

\subsubsection{Azure Quantinuum Cost\cite{AWSQuantumPricing}{\textbf{\textbf{[12]}}}}

Historically, Quantinuum has offered subscription-based plans which allocate H-System Quantum Credits (HQCs) based on the following formula

HQC = 5 + C (Number of 1-qubit gates + Number of 2-qubit gate + 5N) /5000

where:

C = Number of shots

N = Number of state preparation and measurement operations

 Quantinuum has just recently added support for a pay-as-you-go plan. For this project, a monthly subscription was acquired, for \$185K per month, that included 17,000 HQC for either of the Quantinuum machines (H1 and H2) and 170,000 HQC for the emulators. Estimated costs for these credits, for a monthly \$185,000 subscription, were allocated and calculated according to table \ref{tab:quantinuum_cost_allocation}.

\begin{table}[h!]
\caption{Quantinuum hardware / emulation cost allocation}
\centering
\renewcommand{\arraystretch}{1.3}
\begin{tabular}{l@{\hspace{0.3cm}}r@{\hspace{0.3cm}}r@{\hspace{0.3cm}}r@{\hspace{0.3cm}}r}
\hline
 & \textbf{HQC credits} & \textbf{\% Value} & \textbf{Value} & \textbf{\$/credit} \\
\hline
\textbf{Hardware} & 17,000 & 90\% & \$166,500.00 & \$9.7941 \\
\hline
\textbf{Emulator} & 170,000 & 10\% & \$18,500.00 & \$0.1088 \\
\hline
\textbf{Totals} & & 100\% & \$185,000.00 & \\
\hline
\end{tabular}
\label{tab:quantinuum_cost_allocation}
\end{table}

Note that this is an arbitrary allocation of hardware and emulator costs that we made to account for the fact that Quantinuum is the only vendor that currently charges for the use of simulation/emulation.

\subsubsection{AWS Cost\cite{AWSQuantumPricing}{\textbf{\textbf{[13]}}}}

AWS offers both on-demand and reservation-based cost models. The AWS on-demand cost formula is:

(Number of tasks * Per-task Price) + (Number of Shots * Per-shot Price)

 Table \ref{tab:aws_pricing} provides the cost breakdown for the AWS quantum devices used in this study.

\begin{table}[h!]
\caption{AWS on-demand pricing model (per-task, per-shot)}
\centering
\renewcommand{\arraystretch}{1.3}
\begin{tabular}{l@{\hspace{0.3cm}}r@{\hspace{0.3cm}}r}
\hline
 & \textbf{Per-task Price} & \textbf{Per-shot Price} \\
\hline
\textbf{IonQ Aria} & \$0.30 & \$0.03 \\
\textbf{IonQ Forte} & \$0.30 & \$0.08 \\
\textbf{IQM Garnet} & \$0.30 & \$0.00145 \\
\hline
\end{tabular}
\label{tab:aws_pricing}
\end{table}

A cost comparison of running a 10 qubit Quantum Fourier Transform (QFT) algorithm with 500 shots on various hardware via Azure and AWS is provided below in Table \ref{tab:qft_cost_comparison}.

\begin{table}[h!]
\caption{10-qubit / 500 shot QFT cost comparison}
\centering
\renewcommand{\arraystretch}{1.3}
\begin{tabular}{l@{\hspace{0.3cm}}r@{\hspace{0.3cm}}r@{\hspace{0.3cm}}p{6cm}}
\hline
\textbf{Device} & \textbf{Azure} & \textbf{AWS} & \textbf{Notes} \\
\hline
\textbf{IonQ Aria} & \$48.06 & \$15.30 & \\
\textbf{IonQ Forte} & - & \$40.30 & \\
\textbf{IQM Garnet} & - & \$1.03 & \\
\textbf{Quantinuum H Series} & \$2289.86  & - & 233.8 HPC; \$185,000 subscription for 17k HPC. 1 HPC is equivalent to \$9.79 \\
\hline
\end{tabular}
\label{tab:qft_cost_comparison}
\end{table}

\subsection{Target Machine Cost Analysis}

\begin{figure}
    \centering
    \includegraphics[width=0.9\linewidth]{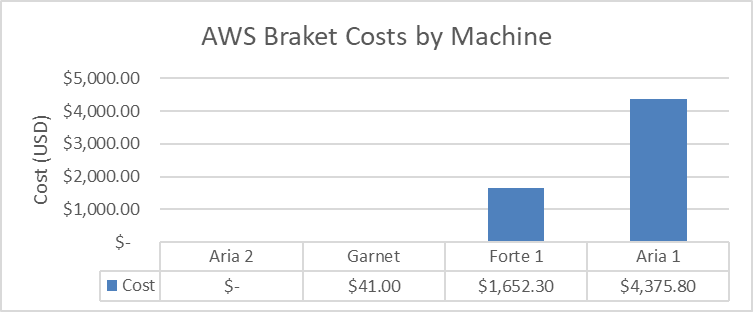}
    \caption{AWS Braket costs by machine }
    \label{fig:AWS_Braket_costs_by_machine}
\end{figure}

\begin{figure}
     \centering
     \includegraphics[width=0.9\linewidth]{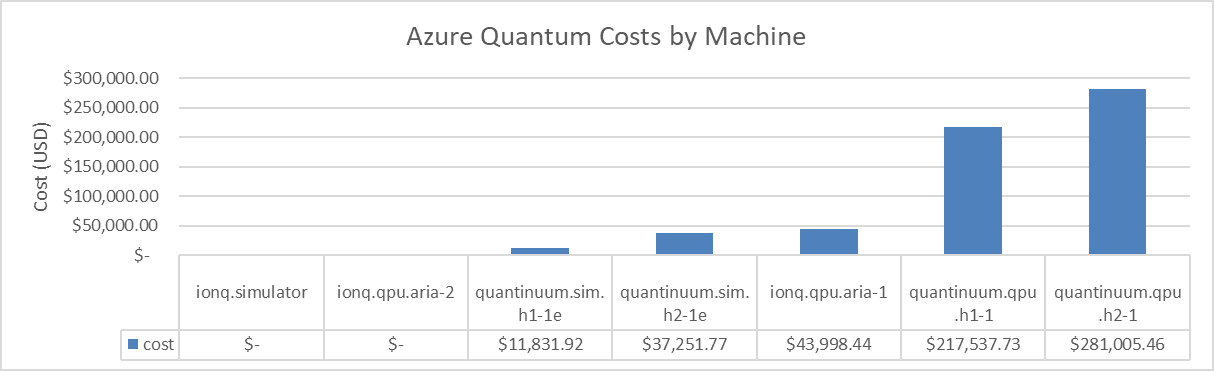}
     \caption{Azure Quantum costs by machine}
     \label{fig:Azure_Quantum_costs_by_machine}
 \end{figure}

\begin{figure}
    \centering
    \includegraphics[width=0.9\linewidth]{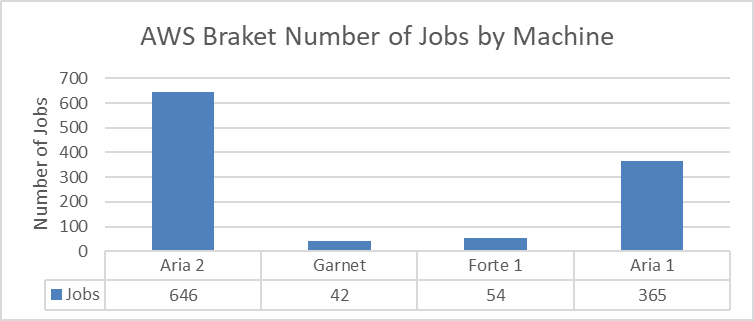}
    \caption{AWS Braket number of jobs by machine}
    \label{fig:AWS_Braket_number_of_jobs_by_machine.png}
\end{figure}

\begin{figure}
     \centering
     \includegraphics[width=0.9\linewidth]{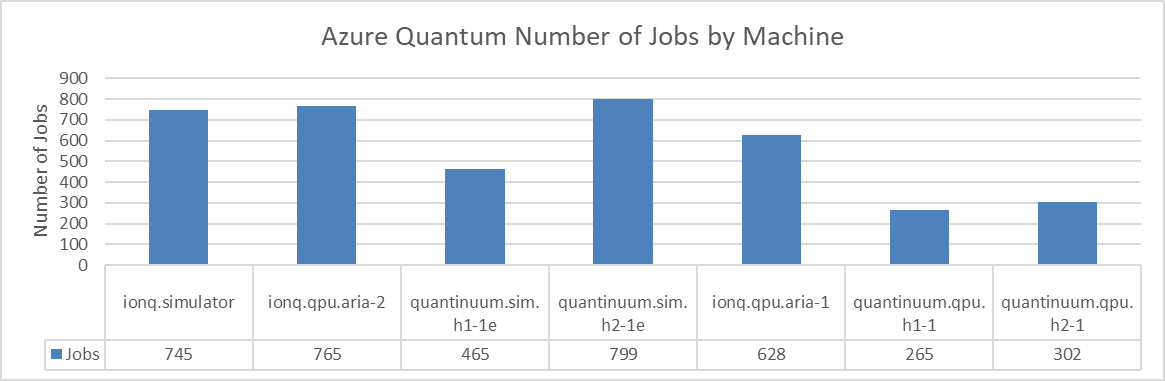}
     \caption{Azure Quantum number of jobs by machine}
     \label{fig:Azure_Quantum_number_of_jobs_by_machine}
 \end{figure}

\begin{figure}
      \centering
      \includegraphics[width=0.9\linewidth]{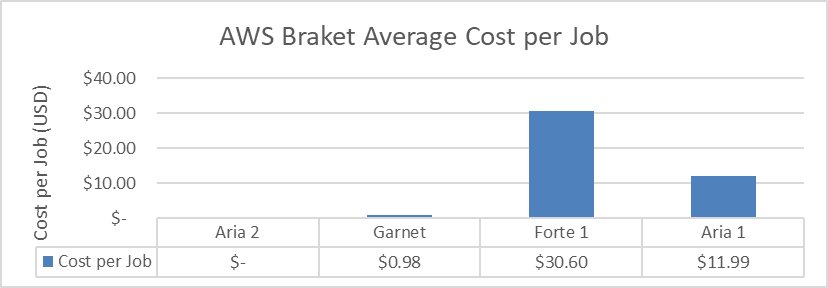}
      \caption{AWS Braket average cost per job}
      \label{fig:AWS_Braket_average_cost_per_job}
  \end{figure}

\begin{figure}
     \centering
     \includegraphics[width=0.90\linewidth]{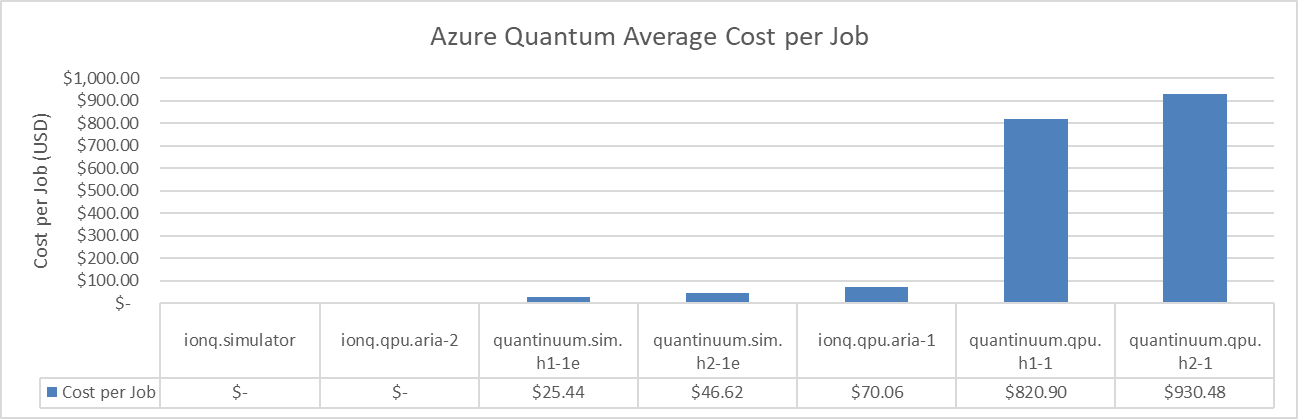}
     \caption{Azure Quantum average cost per job}
     \label{fig:Azure_Quantum_average_cost_per_job}
 \end{figure}

 More than 5,000 jobs were submitted to cloud quantum machines over a 3-month time. 
 The cost of each job was determined and associated with the cloud and machine. 
 The following sections provide this cost data broken down by cloud provider, machine, number of jobs, and average cost per job. 
 The cost breakdowns by cloud provider and machine are shown in 
 Figures \ref{fig:AWS_Braket_costs_by_machine} and \ref{fig:Azure_Quantum_costs_by_machine} for AWS and Azure, respectively. 
 The number of jobs performed by cloud provider and machine are shown in 
 Figures \ref{fig:AWS_Braket_number_of_jobs_by_machine.png} and \ref{fig:Azure_Quantum_number_of_jobs_by_machine}. 
 Finally, the average cost per job, by cloud provider and machine are shown in 
 Figures \ref{fig:AWS_Braket_average_cost_per_job} and \ref{fig:Azure_Quantum_average_cost_per_job}. 
 Note that while there were jobs submitted to Aria 2, the cost was zero since the machine was down for maintenance during the entire data collection period.

\subsubsection{Machine job costs by target and cloud service provider}

\begin{table}[h!]
\caption{Summary of cloud quantum costs}
\centering
\renewcommand{\arraystretch}{1.3}
\begin{tabular}{l@{\hspace{0.3cm}}r@{\hspace{0.3cm}}r}
\hline
\textbf{Machine} & \textbf{Cost} & \textbf{Number of Jobs} \\
\hline
\textbf{AWS Braket} & \textbf{\$6,069.10} & \textbf{1107} \\
\textbf{Aria 2} & \$- & 646 \\
\textbf{Garnet} & \$41.00 & 42 \\
\textbf{Forte 1} & \$1,652.30 & 54 \\
\textbf{Aria 1} & \$4,375.80 & 365 \\
\hline
\textbf{Azure Quantum} & \textbf{\$591,625.33} & \textbf{3969} \\
\textbf{ionq.simulator} & \$- & 745 \\
\textbf{ionq.qpu.aria-2} & \$- & 765 \\
\textbf{quantinuum.sim.h1-1e} & \$11,831.92 & 465 \\
\textbf{quantinuum.sim.h2-1e} & \$37,251.77 & 799 \\
\textbf{ionq.qpu.aria-1} & \$43,998.44 & 628 \\
\textbf{quantinuum.qpu.h1-1} & \$217,537.73 & 265 \\
\textbf{quantinuum.qpu.h2-1} & \$281,005.46 & 302 \\
\hline
\textbf{Grand Total} & \textbf{\$597,694.43} & \textbf{5076} \\
\hline
\end{tabular}
\label{tab:cloud_quantum_costs}
\end{table}

Table \ref{tab:cloud_quantum_costs} shows a summary of the complete cloud quantum costs for this study.

\textbf{The cost of collecting this dataset was dominated by the cost of performing jobs on Quantinuum machines.}

\subsubsection{The cost of fidelity}

\begin{figure}[H]
\includegraphics[width=\textwidth]{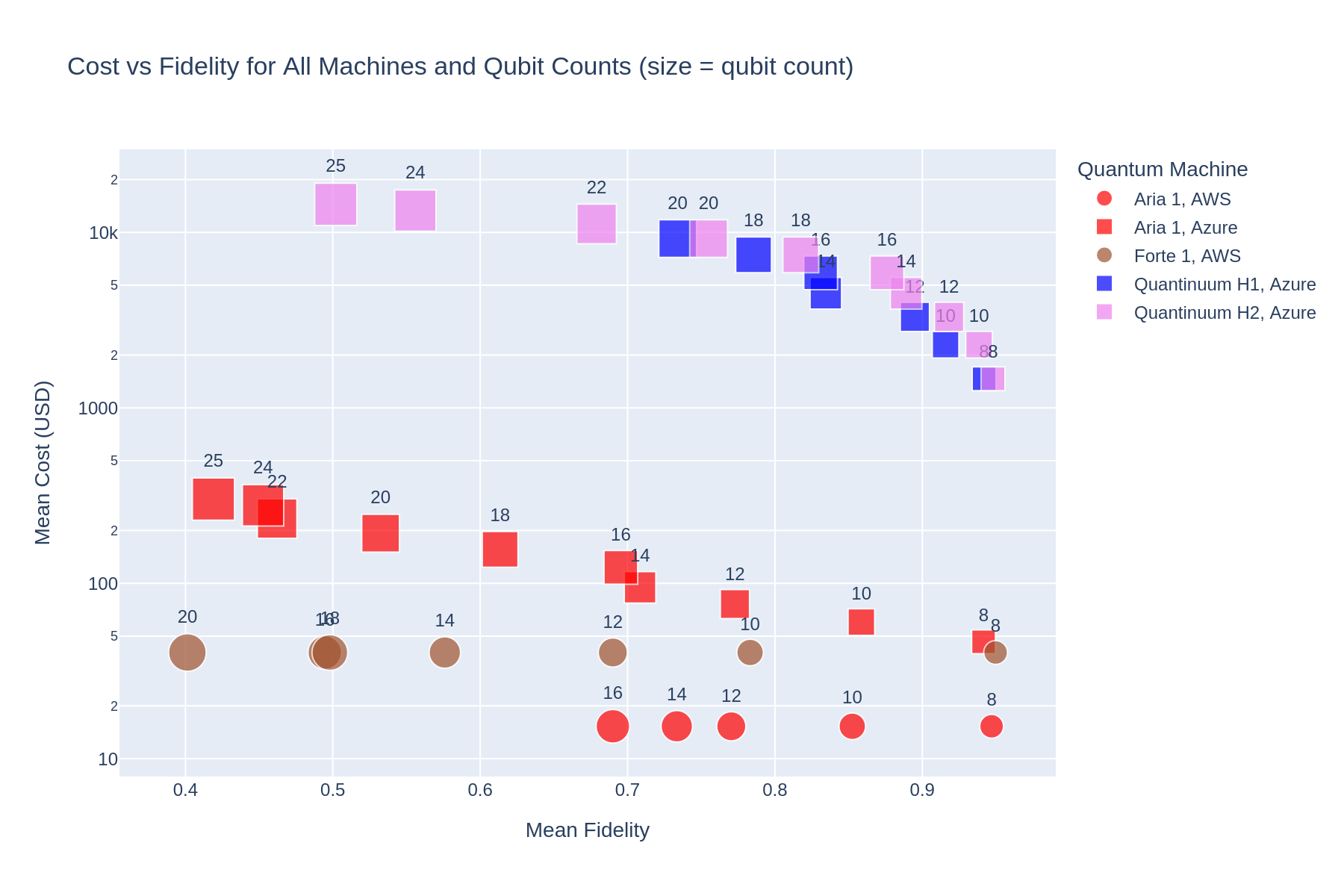}%
\caption{\label{fig:cost_vs_fidelity_machines}The cost of fidelity: job costs as a function of fidelity}
\end{figure}

 The plot in Figure \ref{fig:cost_vs_fidelity_machines} shows the comparative cost of fidelity. Each of the 39 points in this plot shows the mean cost, in USD, versus the mean fidelity, for all the runs, over time, for a given number of qubits (8 to 25), on the same target machine, via the same cloud provider. The size of each symbol is proportional to the number of qubits; each point is also annotated with the number of qubits. The shape of the symbol indicates the cloud provider: square for Azure and circle for AWS. 
 
\begin{figure}[H]
\includegraphics[width=\textwidth]{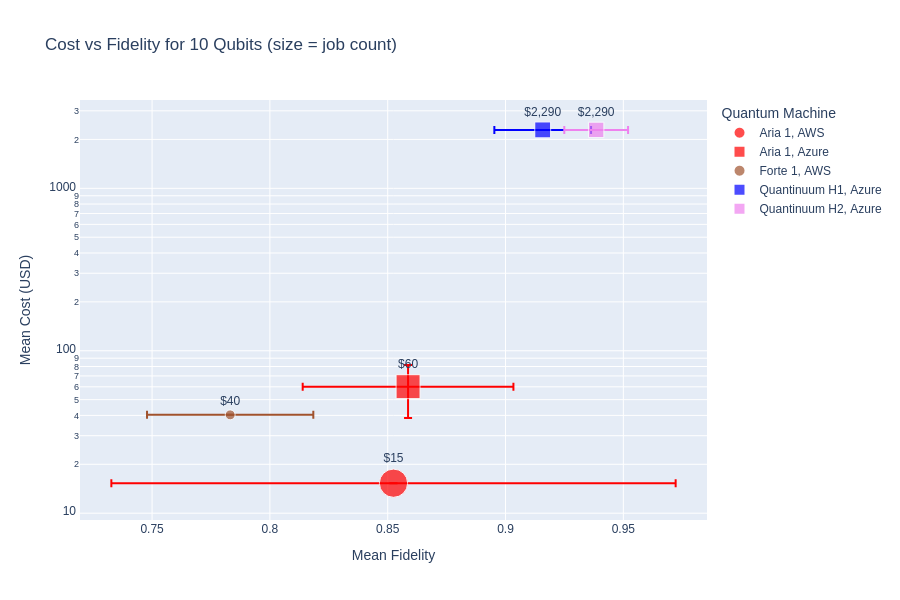}%
\caption{\label{fig:Fidelity_vs_cost_10}The cost of fidelity: job costs as a function of fidelity for 10 qubits}
\end{figure}

\begin{figure}[H]
\includegraphics[width=\textwidth]{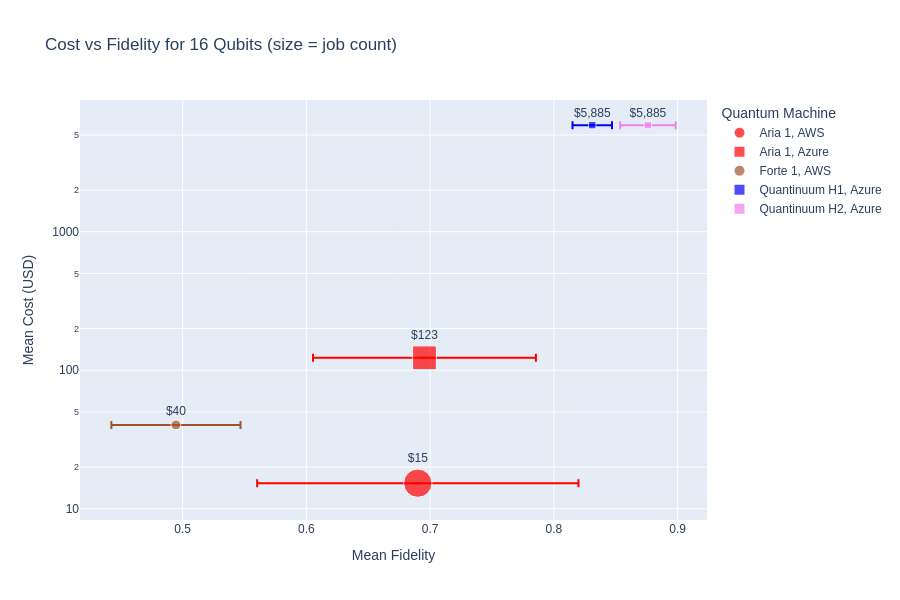}%
\caption{\label{fig:Fidelity_vs_cost_16}The cost of fidelity: job costs as a function of fidelity for 16 qubits}
\end{figure}

\begin{figure}[H]
\includegraphics[width=\textwidth]{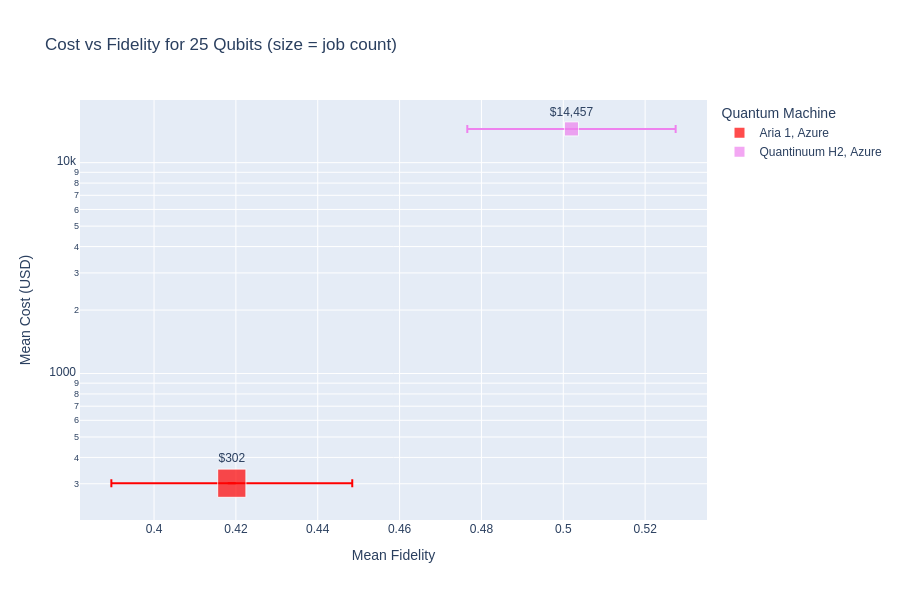}%
\caption{\label{fig:Fidelity_vs_cost_25}The cost of fidelity: job costs as a function of fidelity for 25 qubits}
\end{figure}

 The fidelity generally decreases for a higher number of qubits; the cost typically increases with the number of qubits for Azure but is constant for AWS. 
 For the plots in Figures \ref{fig:cost_vs_fidelity_machines} through \ref{fig:Fidelity_vs_cost_25}, 
 the results are best in the lower-right portion of the plot (lowest cost, highest fidelity); 
 conversely, the results in the upper-left portion of the plot are less desirable (highest cost, lowest fidelity).
 
 The complete 39-point data set can be found in Appendix \ref{app:aggregate_raw_data}.

\begin{adjustwidth}{1cm}{1cm}
\begin{mdframed}[backgroundcolor=lightgray, linecolor=black, nobreak=true]
\textbf{Fidelity disclaimer}: Please note that these runs all used default settings; It is highly likely that experts for these machines could find transpiler and post-processing settings that would produce better fidelity results, particularly for the QFT circuit used for this work. It is not the objective of this work to benchmark the fidelity performance of these machines, but rather to compare default cost and fidelity trade-offs and measure variability over time for given settings.
\end{mdframed}
\end{adjustwidth}

A closer look at these data, for individual numbers of qubits, better reveals the trade-off of fidelity and cost, as well as the expected variability.

Figure \ref{fig:Fidelity_vs_cost_10} shows the data for 10 qubit runs. Note that the Quantinuum machines provide higher fidelity, but at significantly higher cost. The variability of the Quantinuum runs is lower, but this could be affected by the smaller sample size. The results for Aria 1 are significantly less expensive on AWS than for Azure (\$15 vs. \$60), revealing the different cost models between AWS and Azure, while the fidelity is slightly lower and the variability is significantly higher. Recall that the default transpiler settings on AWS are different than for Azure, which is likely the reason for these fidelity and variability differences (after all, these Aria 1 runs are on the exact same machine at IonQ regardless of which cloud provider is used to access them).

We move up to 16 qubits with the data shown in Figure \ref{fig:Fidelity_vs_cost_16}. Notice that cost of Aria 1 on AWS does not change as a function of the number of qubits, due to the AWS cost model. This gives a significant cost advantage to AWS. The Quantinuum results are higher fidelity than either Aria 1 or Forte 1 machines from IonQ, and also significantly more expensive.

Figure \ref{fig:Fidelity_vs_cost_25}, at 25 qubits, shows the largest circuit that we measured. Here we see only Aria 1 and Quantinuum H2 on Azure. Data for Aria 1 on AWS is not available at 25 qubits, as the AWS transpiler settings cause the transpiled circuit to be too large to submit to Aria 1. The number of data points for Quantinuum H2 is small due to the high cost of each run. This plot shows a significant difference in fidelity and the largest difference in cost, showing that, while much more expensive, the default Quantinuum results are measurably better than those found on the Aria 1 machine for this circuit.

\begin{adjustwidth}{1cm}{1cm}
\begin{mdframed}[backgroundcolor=lightgray, linecolor=black, nobreak=true]
\textbf{Cost/Fidelity Trade-offs}: The vast cost range—\$15 to \$14,000—indicates that users must carefully evaluate cost structures and prioritize providers offering the best balance of fidelity and affordability. Our results showed that AWS generally provides the best value, but with limitations of circuit size and results variability. The Quantinuum results were consistently the highest fidelity but at the highest cost. The Azure cost model is more expensive than AWS, but provides access to the highest fidelity Quantinuum machines, which are not available on AWS.
\end{mdframed}
\end{adjustwidth}

\section{Discussion and conclusions}

This study marks a significant milestone in understanding the practical realities of cloud quantum computing, offering a comprehensive, user-focused evaluation of Microsoft Azure Quantum and Amazon AWS Braket over a three-month period from mid-September to mid-December 2024. By executing over 5,000 Quantum Fourier Transform (QFT) jobs on a diverse set of quantum hardware---including IonQ’s Aria and Forte, IQM’s Garnet, and Quantinuum’s H1-1 and H2-1---we have compiled a robust dataset that sheds light on accessibility, reliability, fidelity, and cost. The development of an automated, Python-based software system using the Qiskit SDK and a MongoDB database represents a critical contribution, enabling consistent, reproducible benchmarking across multiple platforms. This system not only facilitated the collection of data for this study but also serves as a scalable framework for future investigations, fostering transparency and extensibility in quantum computing research.

The significance of this work extends beyond its technical findings. By providing a detailed, user-centric perspective, it empowers researchers, developers, and industry practitioners to navigate the complexities of cloud quantum computing, optimizing resource allocation and minimizing costly pitfalls. The study highlights the transformative potential of quantum computing in fields such as cryptography, materials science, and optimization, while also exposing the current limitations of the ecosystem. These insights are crucial for bridging the gap between the theoretical promise of quantum computing and its practical implementation, guiding users in making informed decisions and urging providers to address operational and software challenges.

Our analysis reveals several key findings that encapsulate the state of cloud quantum computing and offer actionable guidance:

\begin{itemize}
    \item \textbf{Transpiler Configuration Critically Impacts Performance}: AWS Braket’s choice for transpiler gate set for IonQ’s Aria 1 resulted in circuits three times larger than those produced by Azure, limiting qubit counts to 16 versus Azure’s 25. This discrepancy, reported to IonQ, underscores the need for users to verify transpiler settings to ensure efficient algorithm execution and maximum hardware utilization.
    \item \textbf{Operational Reliability Remains a Challenge}: Unpredictable queue times, with Azure overestimating wait times in 36\% of cases, and variable machine availability---such as Aria 2’s complete unavailability versus H2-1’s 100\% availability---highlight that cloud quantum computing is not yet a seamless solution. Strategic workflow planning is essential to mitigate disruptions.
    \item \textbf{Fidelity Varies Widely Across Machines and Time}: QFT fidelity ranged from 0.0 to 0.8 for 20 qubits, with IonQ’s simulator initially underestimating fidelity by up to 80\%, while Quantinuum’s emulators closely matched hardware performance (within 1-16\%). This variability emphasizes the importance of reliable simulators for debugging and the need for stable hardware performance.
    \item \textbf{Fidelity/Cost Trade-offs are nonlinear}: Cost increased dramatically for incrementally higher fidelity, from \$15 for IonQ’s Aria on AWS to \$14,000 for Quantinuum on Azure. This difference suggests that users must carefully evaluate pricing models to prioritize providers offering the best balance of performance and affordability.
\end{itemize}

These findings highlight the immaturity of the cloud quantum computing ecosystem, where simple configuration differences, operational inconsistencies, and cost disparities can significantly impact user experience. However, these challenges also present opportunities for improvement. Providers can enhance transpiler optimization, increase queue transparency, stabilize hardware performance, and align pricing with outcomes. Users, meanwhile, can leverage these insights to select platforms strategically, optimize circuit designs, and advocate for better tools and documentation.

The broader impact of this study lies in its role as a catalyst for progress in quantum computing. By exposing operational realities and providing a reusable framework, it paves the way for more reliable, accessible, and cost-effective quantum services. This work not only empowers current users but also accelerates the adoption of quantum technologies across industries, bringing us closer to realizing their revolutionary potential.

\section{Future Work}

Next steps include:
\begin{itemize}
    \item Support for additional services and machines in the provider/machine matrix.
    \item Development of web-based tools to update, summarize, and visualize benchmark performance data as machine offerings evolve.
    \item Investigation of sources of noise and exploration of techniques for modeling/predicting noise in both today’s single QPU machines and future multiple-QPU machines. 
\end{itemize}

\section{Acknowledgments}

The authors acknowledge the Secretary of the Air Force Concepts, Development, and Management organization (SAF/CDM) for their funding and support.

\section{References}

 \bibliography{ThreeMonths}
\newpage
\appendix

\section{Cloud Quantum Computing Aggregate Raw Data}
\label{app:aggregate_raw_data}
The raw data collected for this work included thousands of runs of a QFT circuit on multiple quantum computers, with varying number of qubits, through AWS or Azure cloud services. The resulting fidelity and cost of each run were collected. These runs were performed over approximately a three-month period.  

Table \ref{tab:aggregate_data} shows a summary of these runs. Each row of this table represents all the runs, over time, for a given number of qubits, on the same target machine, via the same cloud provider, for a total of 39 data points. 

The columns represent an index, the number of qubits (Qubits; 8-25), the cloud provider (Cloud; either AWS or Azure), The mean fidelity (Fidelity; 0-1), the standard deviation of the fidelity (Fid. Std), the number of jobs represented by each row (Jobs), the mean cost in USD (Cost), and the standard deviation of cost in USD (Cost Std).  

Target machines were provided by IonQ (Aria 1 and Forte 1) and Quantinuum H1 (Q H1) and Quantinuum H2 (Q H2). Data for IQM machines was collected, but was omitted from this table because all measured fidelity values were 0.0. The Rigetti machines were retired from Azure and thus no data was collected.

\begin{longtable}{@{} l S[table-format=2.0] p{1.5cm} p{2.5cm} S[table-format=1.3] S[table-format=1.3] S[table-format=3.0] S[table-format=5.2] S[table-format=5.2] @{}}
\caption{Quantum Computing Performance Metrics} \label{tab:aggregate_data} \\
\toprule
 & {Qubits} & {Cloud} & {Target} & {Fidelity} & {Fid. Std} & {Jobs} & {Cost (USD)} & {Cost Std} \\
\midrule
\endfirsthead
\caption[]{Quantum Computing Performance Metrics (continued)} \\
\toprule
 & {Qubits} & {Cloud} & {Target} & {Fidelity} & {Fid. Std} & {Jobs} & {Cost} & {Cost Std} \\
\midrule
\endhead
\midrule
\multicolumn{9}{r}{\textit{Continued on next page}} \\
\endfoot
\bottomrule
\endlastfoot
0  & 8  & AWS   & Aria 1      & 0.947 & 0.022 & 56 & 15.30   & 0.00 \\
1  & 8  & AWS   & Forte 1     & 0.950 & 0.012 & 7  & 40.30   & 0.00 \\
2  & 8  & Azure & Aria 1      & 0.942 & 0.024 & 38 & 46.36   & 28.87 \\
3  & 8  & Azure & Q H1        & 0.942 & 0.022 & 19 & 1464.22 & 0.00 \\
4  & 8  & Azure & Q H2        & 0.948 & 0.059 & 17 & 1464.22 & 0.00 \\
5  & 10 & AWS   & Aria 1      & 0.852 & 0.120 & 57 & 15.30   & 0.00 \\
6  & 10 & AWS   & Forte 1     & 0.783 & 0.035 & 7  & 40.30   & 0.00 \\
7  & 10 & Azure & Aria 1      & 0.859 & 0.045 & 41 & 60.11   & 21.50 \\
8  & 10 & Azure & Q H1        & 0.916 & 0.021 & 18 & 2289.86 & 0.00 \\
9  & 10 & Azure & Q H2        & 0.938 & 0.014 & 17 & 2289.86 & 0.00 \\
10 & 12 & AWS   & Aria 1      & 0.770 & 0.152 & 57 & 15.30   & 0.00 \\
11 & 12 & AWS   & Forte 1     & 0.690 & 0.048 & 7  & 40.30   & 0.00 \\
12 & 12 & Azure & Aria 1      & 0.773 & 0.134 & 43 & 76.10   & 12.70 \\
13 & 12 & Azure & Q H1        & 0.895 & 0.022 & 15 & 3296.69 & 0.00 \\
14 & 12 & Azure & Q H2        & 0.918 & 0.013 & 15 & 3296.69 & 0.00 \\
15 & 14 & AWS   & Aria 1      & 0.733 & 0.156 & 58 & 15.30   & 0.00 \\
16 & 14 & AWS   & Forte 1     & 0.576 & 0.065 & 7  & 40.30   & 0.00 \\
17 & 14 & Azure & Aria 1      & 0.708 & 0.140 & 42 & 94.83   & 1.51 \\
18 & 14 & Azure & Q H1        & 0.835 & 0.032 & 4  & 4497.45 & 0.00 \\
19 & 14 & Azure & Q H2        & 0.889 & 0.030 & 4  & 4497.45 & 0.00 \\
20 & 16 & AWS   & Aria 1      & 0.690 & 0.130 & 58 & 15.30   & 0.00 \\
21 & 16 & AWS   & Forte 1     & 0.495 & 0.052 & 7  & 40.30   & 0.00 \\
22 & 16 & Azure & Aria 1      & 0.695 & 0.090 & 41 & 123.04  & 0.00 \\
23 & 16 & Azure & Q H1        & 0.831 & 0.016 & 4  & 5885.27 & 0.00 \\
24 & 16 & Azure & Q H2        & 0.876 & 0.022 & 4  & 5885.27 & 0.00 \\
25 & 18 & AWS   & Forte 1     & 0.498 & 0.026 & 3  & 40.30   & 0.00 \\
26 & 18 & Azure & Aria 1      & 0.614 & 0.112 & 42 & 155.88  & 0.01 \\
27 & 18 & Azure & Q H1        & 0.786 & 0.029 & 4  & 7460.17 & 0.00 \\
28 & 18 & Azure & Q H2        & 0.818 & 0.023 & 4  & 7460.17 & 0.00 \\
29 & 20 & AWS   & Forte 1     & 0.401 & 0.019 & 3  & 40.30   & 0.00 \\
30 & 20 & Azure & Aria 1      & 0.532 & 0.129 & 40 & 192.94  & 0.01 \\
31 & 20 & Azure & Q H1        & 0.734 & 0.027 & 3  & 9226.04 & 0.00 \\
32 & 20 & Azure & Q H2        & 0.755 & 0.030 & 2  & 9226.04 & 0.00 \\
33 & 22 & Azure & Aria 1      & 0.462 & 0.097 & 39 & 233.90  & 0.01 \\
34 & 22 & Azure & Q H2        & 0.679 & 0.027 & 2  & 11179.97 & 0.00 \\
35 & 24 & Azure & Aria 1      & 0.453 & 0.011 & 6  & 278.43  & 0.00 \\
36 & 24 & Azure & Q H2        & 0.556 & 0.003 & 2  & 13319.00 & 0.00 \\
37 & 25 & Azure & Aria 1      & 0.419 & 0.029 & 8  & 301.94  & 0.00 \\
38 & 25 & Azure & Q H2        & 0.502 & 0.025 & 2  & 14457.07 & 0.00 \\
\end{longtable}

\end{document}